\documentclass{osa-article}

\usepackage{upgreek}
\usepackage{pdfpages}

\journal{osajournal}

\articletype{Research Article}

\begin{document}

\title{Observing exceptional point degeneracy of radiation with electrically pumped photonic crystal coupled-nanocavity lasers}

\author{Kenta Takata,\authormark{1,2,*} Kengo Nozaki,\authormark{1,2} Eiichi Kuramochi,\authormark{1,2} Shinji Matsuo,\authormark{1,3} Koji Takeda,\authormark{1,3} Takuro Fujii,\authormark{1,3} Shota Kita,\authormark{1,2} Akihiko Shinya,\authormark{1,2} Masaya Notomi,\authormark{1,2,4,5}}

\address{\authormark{1}Nanophotonics Center, NTT Corporation, 3-1 Morinosato-Wakamiya, Atsugi, Kanagawa 243-0198, Japan\\
\authormark{2}NTT Basic Research Laboratories, NTT Corporation, 3-1 Morinosato-Wakamiya, Atsugi, Kanagawa 243-0198, Japan\\
\authormark{3}NTT Device Technology Laboratories, NTT Corporation, 3-1 Morinosato-Wakamiya, Atsugi, Kanagawa 243-0198, Japan\\
\authormark{4}Tokyo Institute of Technology, 2-12-1 Ookayama, Meguro-ku, Tokyo 152-8550, Japan}

\email{\authormark{*}kenta.takata.ke@hco.ntt.co.jp} 
\email{\authormark{5}masaya.notomi.mn@hco.ntt.co.jp}



\begin{abstract}
Controlling gain and loss of coupled optical cavities can induce non-Hermitian degeneracies of eigenstates, called exceptional points (EPs). Various unconventional phenomena around EPs have been reported, and expected to incorporate extra functionalities into photonic devices. The eigenmode exactly under the EP degeneracy is also predicted to exhibit enhanced radiation. However, such responses have yet to be observed in on-chip lasers, because of both the limited controllability of their gain and loss and the lifting of degeneracy by pump-induced cavity detuning. Here, we report the first non-Hermitian nanophotonic platform based on two electrically pumped photonic crystal lasers and its spontaneous emission at an EP degeneracy. Systematically tuned and independent current injection to our wavelength-scale active heterostructure cavities enables us to demonstrate the clear EP phase transition of their spontaneous emission, accompanied with the spectral coalescence of coupled modes and reversed pump dependence of the intensity. Furthermore, we find experimentally and confirm theoretically the peculiar squared Lorentzian emission spectrum very near the exact EP, which indicates the four-fold enhancement of the photonic local density of states induced purely by the degeneracy. Our results open a new pathway to engineer the light-matter interaction by non-Hermiticity and explore larger reconfigurable laser arrays for further non-Hermitian features and physics.
\end{abstract}

\section{Introduction}
Coupled optical cavities and waveguides with imaginary refractive index contrast, i.e. distributed gain and loss, can exhibit peculiar degeneracies called exceptional points \cite{Kato2013,Heiss2012,Feng2017non,ElGanainy2018non,Ozdemir2019par} (EPs). In such a system, eigenmodes undergo a transition between two phases which are divided by the EP. One phase comprises extended supermodes with parity-time (PT) symmetry \cite{Bender1998,Makris2008,Guo2009,Ruter2010}. Here, the real parts of their eigenfrequencies and propagation constants are split. In contrast, the imaginary parts of them are clamped at the average of the imaginary effective potential, canceling its local contribution over the unit cell (symmetric phase). In the other regime, PT symmetry is spontaneously broken; the eigenstates localize at either the amplifying or de-amplifying elements (broken phase). Correspondingly, the split real spectrum coalesces at the EP, and then the imaginary spectrum bifurcates into two or more branches, with singular a dependence on parameters involved. This EP transition leads to intriguing features, such as reversed pump dependence \cite{Liertzer2012,Brandstetter2014,Peng2014los}, single-mode oscillation \cite{Feng2014sin,Hodaei2014par}, and enhanced sensitivity \cite{Hodaei2017enh,Chen2017exc}.

There has also been rising interest in the photonic EP degeneracy itself. Distinct from the accidental degeneracy of characteristic eigenvalues in Hermitian systems with orthogonal modes, the EP makes not only some eigenvalues but also corresponding eigenmodes identical. Thus, the effective non-Hermitian Hamiltonian becomes non-diagonalizable. The resultant nonorthogonal eigenstates surrounding the EP can enjoy optical isolation \cite{Peng2014par,Chang2014par}, coherent absorption \cite{Longhi2010,Wong2016}, unidirectional reflectivity \cite{Lin2011,Regensburger2012,Feng2013exp}, and asymmetric mode conversion \cite{Doppler2016,Yoon2018}. 

Although many papers have studied phenomena around the EP, observing optical responses at the EP degeneracy has been a persistent technical challenge, even for basic two-cavity devices \cite{Brandstetter2014,Peng2014los,Hodaei2014par,Peng2014par,Chang2014par,Zhu2018,Gao2019non,Kim2016,Gao2017,Liu2017,Yao2019,Hayenga2019}. In fact, the EP degeneracy is predicted to have significant influence on radiation processes \cite{Yoo2011,Lin2016,Pick2017gen,Pick2017enh}. However, it is a single spot in the continuous parameter space for eigenfrequencies; therefore, fine and independent control of gain and loss is required for each cavity, which is demanding for systems based on passive loss processes or optical pumping. To this end, preparing strongly coupled lasers with current injection is desirable. Meanwhile, carrier plasma and thermo-optic effects arising with asymmetric pumping induce detuning of their resonance frequencies. This active mismatch lifts directly the degeneracy of the eigenfrequencies \cite{Brandstetter2014,Gao2017}. In addition, it results in the significant damping of one of the coupled modes \cite{Zhu2018,Gao2019non,Kim2016}, which hampers their coalescence and hence the EP response. Multiple cavity modes with comparative $Q$ factors \cite{Liu2017,Yao2019,Hayenga2019} are also subject to carrier-mediated mode competition that can disrupt the pristine properties at the EP.

Here, we report the observation of spontaneous emission under the EP degeneracy with two current-injected photonic crystal lasers. We establish the first nanocavity-based non-Hermitian platform with electrical pumping, by using our buried heterostructure technique \cite{Matsuo2010,Takeda2013,Takata2017}. It is generally hard to achieve lasing in electrically pumped nanocavities (i.e. cavities with wavelength- or subwavelength-scale mode volumes), because of restricted gain and difficulty in thermal management. Thus, the wavelength-scale active heterostructure with photonic crystals \cite{Takeda2013} operates as the only current-driven continuous-wave nanocavity laser at room temperature, in the present conditions. Now, we successfully integrate two of them with strong coupling, which also hold continuous-wave room-temperature oscillation, and explore the exact EP response of their emission. Efficient carrier injection and high heat conductivity in the tiny heterostructures enable minimal pump-induced resonance shift and stable control of gain and loss for each cavity. Selective high $Q$ factors for their coupled ground modes are also achieved so that the mode competition is suppressed. We first investigate the system with highly asymmetric pumping. Here, we clarify that, whenever there is non-negligible cavity detuning, it is barely possible for the lasing PT-symmetric supermodes to reach any degree of non-Hermitian coalescence. In contrast, our elaborate measurement and analysis of the spontaneous emission demonstrate the distinct EP transition without severe detrimental effects, and identify the fine EP location. Remarkably, we find the squared Lorentzian emission spectrum very near the exact EP, which signifies the unconventional enhancement of the photonic local density of states (LDOS) \cite{Yoo2011,Lin2016,Pick2017gen,Pick2017enh,Wijnands1997}. Our results provide a new approach to handle the light-matter interaction and light emission.

\section{Theoretical backgrounds}
We consider two identically designed optical cavities with spatial proximity and imaginary potential contrast [Fig. 1(a)]. Their ground cavity modes exchange photons with evanescent waves, and thus the system eigenfrequencies are split by the mode coupling, $\kappa$. However, the gain and loss can counteract the frequency splitting by the EP transition. The first-order temporal coupled mode equations (CMEs) \cite{Yariv1999} for the complex cavity-mode amplitudes $\left\{a_i(t)\right\}$ are derived as,
\begin{equation}
\frac{d}{dt} \Bigg(\begin{array}{c}
	a_1\\
	a_2
\end{array}\Bigg)
=\Bigg(\begin{array}{cc}
	i(\omega_0+\delta)-\gamma_1 & i\kappa\\
	i\kappa & i(\omega_0-\delta)-\gamma_2
\end{array}\Bigg) \Bigg(\begin{array}{c}
	a_1\\
	a_2
\end{array}\Bigg),
\end{equation}
where $\gamma_i$ is the loss (positive) or gain (negative) for each cavity, and $\omega_0$ is the average resonance frequency. Without loss of generality, small cavity detuning to $\omega_0$ is introduced as $\pm\delta\in\mathbb{R}$. The model reduces to the eigenvalue problem with the ansatz $(a_1, \ a_2)^{\rm T} = (A_1, \ A_2 )^{\rm T} e^{i \omega t}$. The resultant eigen-detuning, $\Delta \omega_i \equiv \omega - \omega_0 = i (\gamma_1 + \gamma_2)/2 \pm \sqrt{\kappa^2 - [(\gamma_1 - \gamma_2)/2 - i \delta ]^2}$, turns into an EP when its second term vanishes: $\gamma_1 - \gamma_2 = 2\kappa$, $\delta = 0$ [Fig. 1 (b)]. Here, the two eigenmodes become degenerate and chiral, $(A_1, \ A_2 )^{\rm T} = (1, \ -i)^{\rm T}/\sqrt{2}$. Meanwhile, it is notable that $\delta$ generally resolves the exact degeneracy and smooths the nearby singular spectrum.
\begin{figure}[t!]
	\centering\includegraphics[width=13cm]{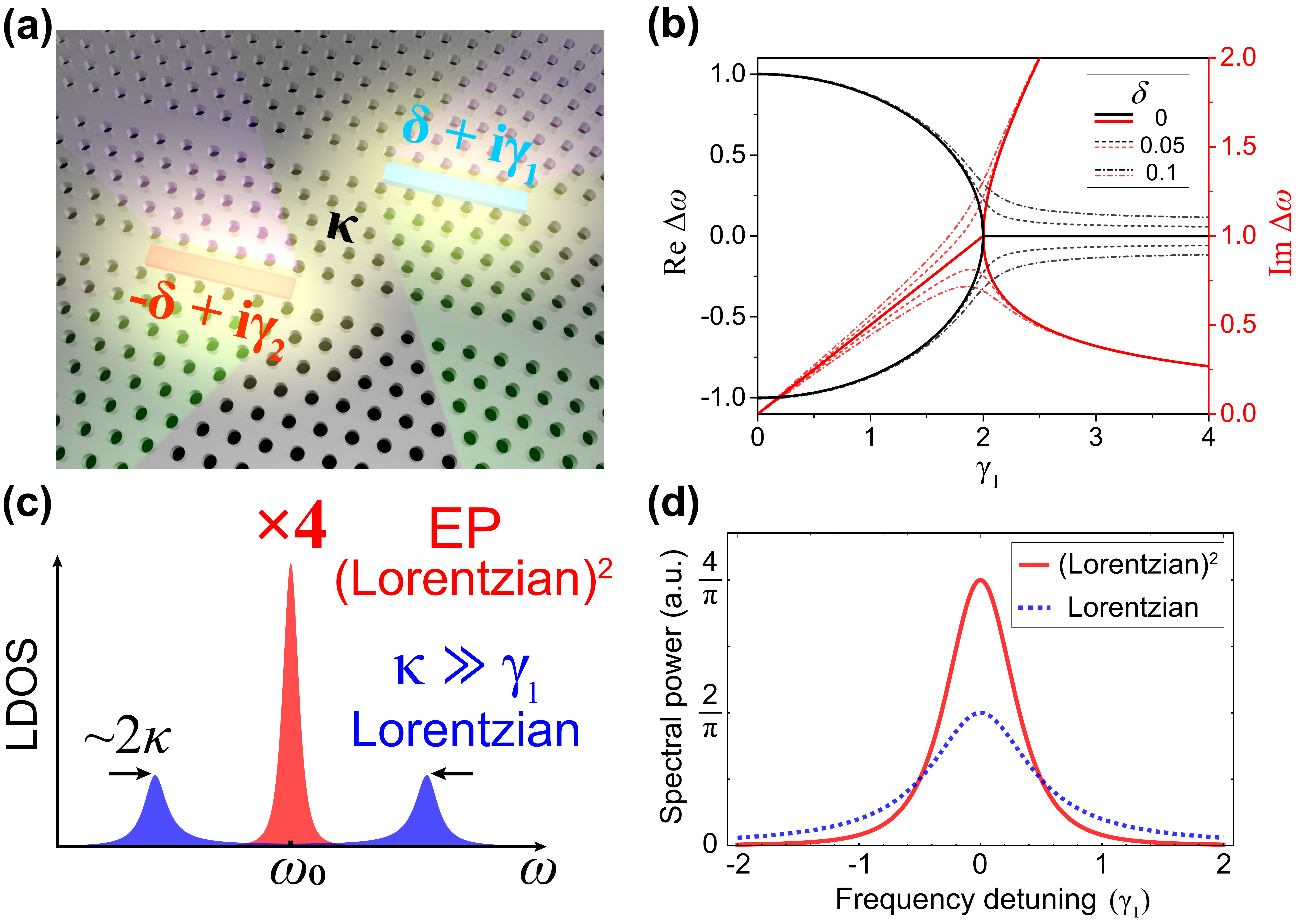}
	\caption{Spontaneous emission of two coupled non-Hermitian nanolasers. (a) Schematic of the system. Cavity $i$ has frequency detuning $(-1)^{i-1} \delta$ to their average resonance $\omega_0$, local loss $\gamma_i$, and an evanescent coupling $\kappa$ with the other. (b) EP transition of the complex eigen-detuning $\Delta \omega_i = \omega_i - \omega_0$ in reference to the coupling, $\kappa=1$, for $\gamma_2 = 0$. The EP is at $(\gamma_1, \delta) = (2, 0)$, and finite $\delta$ blurs the sharp coalescence of the two branches. (c) Comparison of photonic LDOS for the system in the large coupling limit ($\kappa \gg \gamma_1$) and that at the EP ($2\kappa = \gamma_1$) for $\gamma_2 = 0$. The spectral LDOS of the EP resonance has a squared Lorentzian shape, and its peak is four times higher than that for one of the split Lorentzian supermodes far from the EP. (d) Lorentzian and squared Lorentzian spectral functions based on the same loss factors $(\gamma_1 > 0, \gamma_2 = 0)$ and integrated intensity. The EP degeneracy doubles the peak power, compared to the sum of two orthogonal Lorentzian modes with a linewidth of $\gamma_1$ (Hermitian diabolic point).}
\end{figure}

The EP exhibits peculiar radiation responses \cite{Yoo2011,Lin2016,Pick2017gen,Pick2017enh} [Fig. 1(c)]. When the system is in the symmetric phase and the eigenfrequency splitting is large, the spectral LDOS \cite{Wijnands1997} of the two coupled modes are Lorentzian functions with ideally the same linewidth. At the non-Hermitian degeneracy, however, the two spectral peaks coalesce and constructively interfere with each other. Thus, the resultant radiation power spectrum, which is directly relevant to the LDOS, takes on a squared Lorentzian shape. When the system has a transparent cavity and a lossy cavity (e.g. $\gamma_1 > 0, \gamma_2 = 0$), the corresponding peak LDOS is increased purely by the effect of the degeneracy. Such enhancement at this {\it passive} EP is four-fold, compared with each of the separate peaks in the large coupling limit. Namely, compared to the mere sum of the two Lorentzian modes (i.e. Hermitian accidental degeneracy of two orthogonal states), the EP resonance with the common loss and the same integral intensity has a doubly high peak and $\sqrt{\sqrt{2} - 1} \approx 0.644$ times narrower linewidth [Fig. 1(d); see also Section 9 of Supplement 1]. Active cavities with spontaneous emission, i.e. flat spectral excitation via the pumped gain media, are well suited for its demonstration. In contrast, EPs with nonlinear processes can generate excess noise and result in their linewidth broadening \cite{Zhang2018}.

\section{Experimental set-up and EP transition in lasing regime}
We prepared a sample comprising two coupled photonic crystal lasers based on buried heterostructure nanocavities \cite{Matsuo2010,Takeda2013,Takata2017} [Fig. 2(a); see also Section 1 of Supplement 1]. Here, gain media with six quantum wells (colored red), which work as mode-gap cavities, are embedded in an air-suspended InP photonic crystal slab. Two line defects narrower than the lattice-matched width improve the cold $Q$ factors of the coupled ground cavity modes. DC current is applied and controlled for each cavity via independent PIN junctions. Note that a single-laser device with a commensurate electric channel has a low lasing threshold $I_{\rm th}$ of about $37 \ \upmu {\rm A}$, at which it has a high $Q$ factor of 14,000 (Section 2 of Supplement 1). When symmetrically pumped below the threshold with $30 \ \upmu {\rm A}$ for comparison, the two-laser sample gives spontaneous emission of the two coupled modes with $Q = 4,000$, slightly below that of the single diode (5,000). Their resonance peaks with a splitting of about 1.0 nm in reference to 1529.7 nm indicates $\kappa = 61 \ {\rm GHz}$ (Section 4 of Supplement 1). It agrees well with the coupling of the simulated ground modes, $\kappa_{\rm sim} = 65 \ {\rm GHz}$ [Fig. 2(b) and Section 1 of Supplement 1]. The near-field emission from both lasers is also observed [Fig. 2(b), inset]. 

We fix the injection current $I_2$ for channel 2 on the left and sweep that to the right, $I_1$ for channel 1, to vary the imaginary potential contrast $\gamma_1 - \gamma_2$. As a result, the detected ground-mode power systematically recovers by the reduction of the local current $I_1$ [Fig. 2(c)]. This reversed pump dependence \cite{Liertzer2012,Brandstetter2014} indicates the EP transition (see also Section 1 and 3 of Supplement 1). Heavy pumping $I_2 = 800 \ \upmu {\rm A}$ maximizes the ratio $P_{\rm R}/P_{\rm min}$ between the power $P_{\rm R}$ for zero bias along channel 1 ($V_1 = 0$) and the minimum value $P_{\rm min}$ in terms of $I_1$. Here, cavity 2 provides gain for achieving notable loss-induced revival of lasing \cite{Peng2014los}. However, the system is critically affected by the cavity detuning $\delta$ and hence misses the EP degeneracy.

Fig. 2(d) depicts the device emission spectra in the lasing regime for constant $I_2 = 800\ \upmu {\rm A}$ and different $I_1$, measured with an optical spectrum analyzer. Here, some leakage current from channel 2 induces a negative $I_1 \approx - 6 \ \upmu {\rm A}$ for $V_1 = 0$. However, the data and hence loss $\gamma_1$ in cavity 1 consistently change under the reverse current. As $I_1$ decreases from $I_1 = 100 \ \upmu {\rm A}$ and $\gamma_1$ hence increases, the blue-side peak $|\lambda_{-} \rangle$ damps, while the other red-side one $|\lambda_{+} \rangle$ remains bright. This is a direct reflection of finite detuning $\delta$, with which the asymmetric pumping $I_2 \gg I_1$ selectively excites the coupled mode closer to the solitary resonance of cavity 2, $\omega_{0} - \delta$. Eventually, the power of $|\lambda_{+} \rangle$ also drops sharply around $I_1=5.4 \ \upmu {\rm A}$, indicating the suppression of oscillation. However, it is $|\lambda_{-} \rangle$ that undergoes the revival of lasing, accompanied with a kinked rise in power and linewidth narrowing (Section 12 of Supplement 1). Such switching of the dominant mode has been observed in relevant studies \cite{Kim2016,Assawaworrarit2017} and attributed to the pump-induced sign flip of $\delta$. The restored peak moves toward the middle of the original coupled-mode resonances by further reducing $I_1$ and hence evidences the EP transition in our device. The near-field patterns for selected $I_1$ [Fig. 2 (e)] not only show the above-mentioned processes in the real space but also exhibit clear mode localization at cavity 2 in the intensity recovery, supporting the PT symmetry breaking.
\begin{figure}[t!]
	\centering\includegraphics[width=12.8cm]{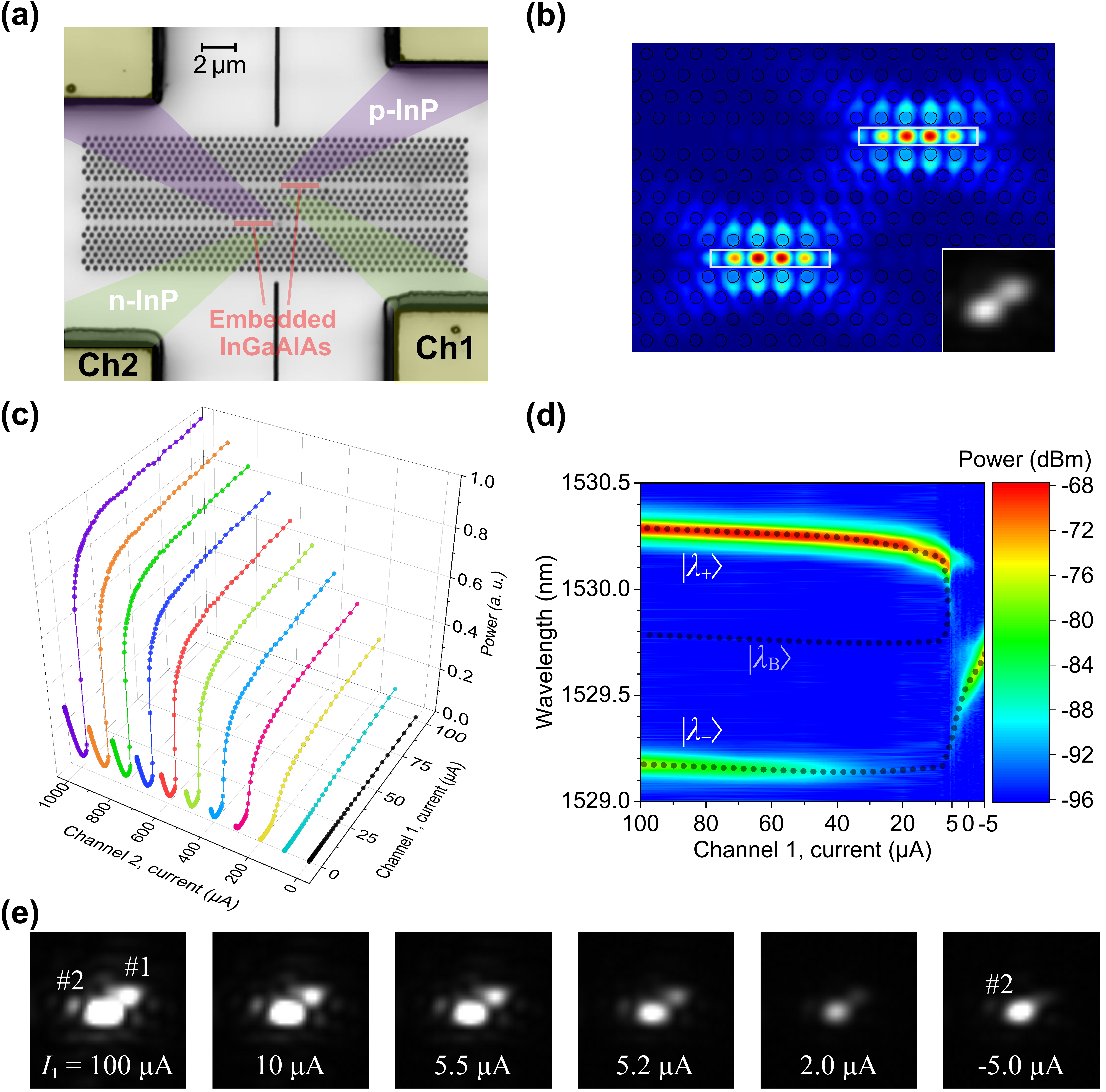}
	\caption{Electrically pumped photonic crystal lasers and EP transition of their lasing ground modes. (a) False-color laser microscope image of the sample. Each of the two buried InGaAlAs heterostructure nanocavities (red squares) has six quantum wells. Diagonally patterned doping layers (purple: p-doped, green: n-doped) with contact pads (yellow) provide independent electric channels for the cavities (right and left: channel 1 and 2 with current $I_1$ and $I_2$, respectively). (b) Magnetic field of a ground supermode for $R = 104.4 \ {\rm nm}$ simulated by the finite-element method. The theoretical cavity coupling is $\kappa_{\rm sim} \approx 65 \ {\rm GHz}$ and the eigenmodes' wavelengths are 1529.25 and 1530.27 nm. Inset: near-field image of the device emission for $I_1=I_2=100\ \upmu {\rm A}$. (c) Current-in and light-out (I-L) curves of the filtered ground modes' emission for several fixed $I_2$ values and swept $I_1$. The photodetector signal shows the systematic reversed pump dependence of the coherent device emission when $I_1$ is small. (d) Device emission spectra for fixed $I_2=800\ \upmu {\rm A}$ and varied $I_1$, measured with an optical spectrum analyzer. As $I_1$ drops, the spectrum features continuous decay of the lower branch $|\lambda_{-} \rangle$, followed by a discrete suppression of the upper branch $|\lambda_{+} \rangle$ and the revival of $|\lambda_{-} \rangle$. Black dots: eigenvalue fitting with Ref. \cite{Assawaworrarit2017}. (e), Corresponding near-field patterns for different $I_1$, showing clear localization of emission at cavity 2 with decreasing $I_1$. The abruptly darkened signal from $I_1=5.5$ to $2.0 \ \upmu {\rm A}$ and glare spot with $I_1 = -5.0 \ \upmu {\rm A}$ support the suppression and revival of lasing by the transition.}
\end{figure}

The steady oscillation condition ${\rm Im} \ \omega_{\rm e} = 0$ enables us to estimate the eigenfrequencies $\omega_{\rm e}$ for the lasing spectra \cite{Assawaworrarit2017}, despite that the system here provides an adaptive (variable) gain $\gamma_{2} < 0$ (Section 10 of Supplement 1). By considering an average effect of detuning $2 \delta = -14.1 \ {\rm GHz}$ and additional thermal and carrier shifts, our numerical analysis (black dots) shown in Fig. 2(d) successfully explains the major portion of the experimental data. Remarkably, one of the eigenmodes manifests itself as two different branches that correspond to different $\gamma_2$. One is the observable coupled mode $|\lambda_{+} \rangle$ in the symmetric phase. The other is the virtual middle branch $|\lambda_{\rm B} \rangle$, which is the same eigenstate in the broken phase, requires larger gain, and still satisfies $\omega_{\rm e} \in \mathbb{R}$. They are annihilated as a pair with a singularity, which does not represent an EP, and turn into a damping mode (${\rm Im} \ \omega_{\rm e} \neq 0$). This destabilization always occurs for finite cavity detuning $\delta$, before the system obtains the loss $\gamma_{1,{\rm EP}} = \kappa$ necessary to reach the only EP in oscillation with $\delta = 0$. Our analysis hence means that it is infeasible for lasing coupled modes to be coalesced by gain and loss, as long as $\delta$ is larger than their narrow linewidths. This is why the EP transition with just a single mode is mostly observed in lasing systems \cite{Brandstetter2014,Gao2019non,Kim2016,Gao2017}, including our result here with revived $|\lambda_{-} \rangle$. Note that $|\lambda_{+} \rangle$ in experiment actually splits into two subpeaks, and one remaining around 1530.15 nm is attributed to an unstable (non-steady) state \cite{Kominis2017}. Additional data are shown in Section 11 of Supplement 1.

\section{EP degeneracy of spontaneous emission}
The spontaneous emission (non-lasing) regime, in contrast, enables us to observe a clear EP transition with spectral coalescence of the two coupled ground modes, as shown in Fig. 3(a) for $I_2=100 \ \upmu {\rm A}$ and decreasing $I_1$. Here, the oscillation threshold for the case of pumping only one of them is about $200 \ \upmu {\rm A}$, because the other cavity and its doped layers behave as additional absorbers (Fig. S2 of Supplement 1). The radiation was measured by a spectrometer with a cryogenic InGaAs line detector (see Section 1 of Supplement 1). In Fig. 3(a), the two distinct spectral peaks originally at 1529.3 and 1530.2 nm coalesce when $I_1 \approx 2 \ \upmu {\rm A}$. In addition, the peak count of the merged resonance at 1529.9 nm increases back to the saturation level of about 55,000 for $I_1 = 0$, confirming the reversed pump dependence (Section 5 of Supplement 1). Although weak higher-order modes are also found around 1523.4 nm [bottom of Fig. 3(a)] and 1521.5 nm (not shown), they are hardly affected by the change in $I_1$. This means that the mode competition is insignificant, because the ground modes have $Q$ factors sufficiently higher than those of other modes. We emphasize that the eigenmodes observed here do not lase and are hence in the spontaneous emission regime, because their spectral linewidths (0.40 nm at least) are fairly broader than that for the single cavity on the lasing threshold (0.11 nm, Fig. S1). Since this is true of both the low-loss coupled modes for $I_1 = 8 \ \upmu {\rm A}$, $I_2=100 \ \upmu {\rm A}$ and the localized mode for $I_1 = 0$, $I_2=100 \ \upmu {\rm A}$ in the broken phase, cavity 2 does not provide notable gain, i.e. $\gamma_2 \approx 0$, despite $I_2 > I_{\rm th}$ for the single laser.

To analyze the system response theoretically, we performed the Fourier transform of the CMEs [Eq. (1)] for the spectral cavity amplitudes $a_i(\omega) = \mathcal{F}[a_i (t)] = a_i (t) e^{-i \omega t }dt$, together with net cavity excitation fields $\left \{ c_i(\omega) \right \}$ arising from the pumping. Because $I_2$ is sufficiently larger than $I_1$ over the entire measurement, we neglect the excitation of cavity 1 for simplicity, $c_1 = 0$. By solving the resultant linear equation (shown in Section 1 of Supplement 1), we reach,
\begin{equation}
\Bigg(\begin{array}{c}
	a_1(\omega)\\
	a_2(\omega)
\end{array}\Bigg)
=\frac{c_2(\omega)}{\kappa^2 + \left[\gamma_1 + i(\Delta\omega - \delta) \right] \left[\gamma_2 + i(\Delta\omega + \delta) \right]}
\Bigg(\begin{array}{c}
	i\kappa\\
	\gamma_1 + i(\Delta\omega - \delta)
\end{array}\Bigg).
\end{equation}
Here, we assume that the spontaneous emission from the medium of cavity 2 has an ideally flat (white) spectrum, $|c_2(\omega)|^2 = {\rm const.}$, because the ground modes ($\approx 1530 {\rm nm}$) are located in the long tail of the heterostructures' luminescence spectrum peaked near 1440 nm. As a result, the spectral intensity of the indirectly pumped cavity $\left| a_1(\omega) \right|^2$ reflects directly the LDOS of the system, which was derived from a singular perturbation analysis \cite{Pick2017gen}. Note that the spectral shape of $\left|a_2(\omega)\right|^2$ is additionally but slightly affected by the relative resonance of cavity 1, i.e. $\Delta\omega - \delta$ on the numerator in Eq. (2) (see also Section 1 and 9 of Supplement 1).
\begin{figure}[t!]
	\centering\includegraphics[width=13cm]{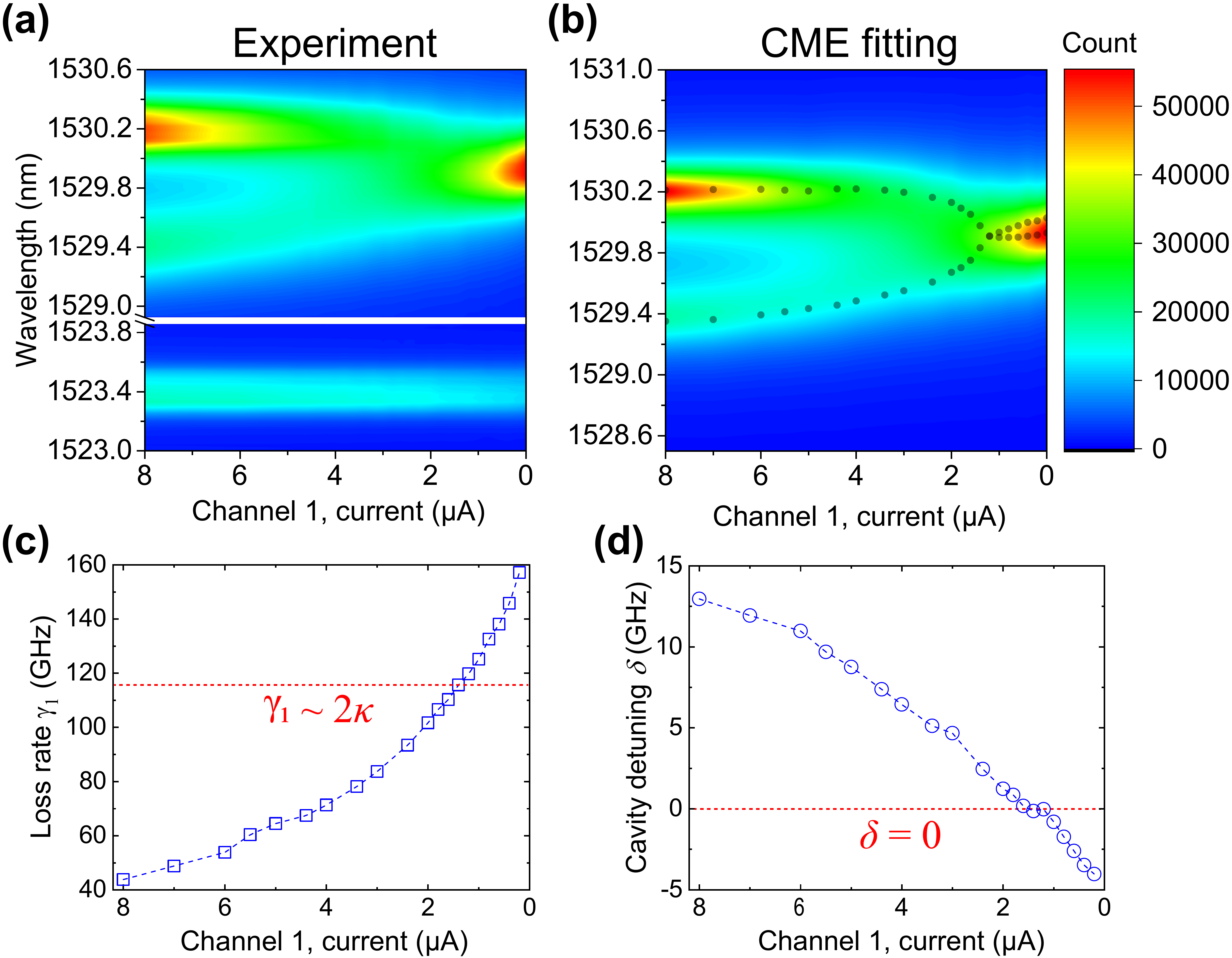}
	\caption{Spectroscopy of the EP transition in the sample's spontaneous emission. (a) Color plot of the observed spectra for constant $I_2 = 100\ \upmu {\rm A}$ and varied $I_1$. Upper: coupled ground eigenmodes, clearly exhibiting the EP transition with the spectral peak coalescence and reversed pump dependence of the peak intensity. Lower: most visible higher order mode, which is hardly affected by the ground-mode process. (b) Result of theoretical fitting via the coupled-mode analysis. It reproduces the experimental spectra well and enables parameter estimation within the model. Here, $\kappa \approx 58 \ {\rm GHz}$ over the entire analysis. Black dots: eigen-wavelengths calculated with the obtained parameters, including their nearly exact coalescence. (c), (d) Estimated (c) loss rate $\gamma_1$ of cavity 1 and (d) cavity detuning $\delta$, dependent on $I_1$. The EP condition for $\gamma_1$ is $\gamma_1 = 2 \kappa \approx 116 \ {\rm GHz}$, and the corresponding closest measurement point is $I_1=1.4 \ \upmu {\rm A}$. Note that the detrimental detuning is almost cancelled out there, by the suppression of the carrier plasma effect with decreasing $I_1$.}
\end{figure}

The theoretical fitting for the spectral data involves the detailed conditions of the optical collection system. Because $a_1$ and $a_2$ hold phase coherence with evanescent coupling, their radiation is expected to have a spatial (directional) intensity distribution due to interference \cite{Gao2017}. The detector signal hence depends on the position of the objective lens controlled by the three-axis nano-positioner. Here, it is aligned so that the out-coupled intensity at the coalescence is maximized. Considering that the degenerate eigenstate is $(1, \ -i)^{\rm T}/\sqrt{2}$, we take the analytic power spectrum for our measurement as $P(\omega) = \eta \gamma_{\rm cav} \left| a_1(\omega) + i a_2(\omega) \right|^2$, under the premise that the identically designed cavity modes have the same radiation loss $\gamma_{\rm cav}$ and collection efficiency $\eta$. Note that our I-L data assure that the system detects the light from both cavity 1 and 2 [Fig. 2(c) and Section 3 of Supplement 1]. For other major possibilities like $\eta \gamma_{\rm cav} \left| a_1(\omega) \pm a_2(\omega) \right|^2$, one of the coupled modes is cancelled out in the symmetric phase, and the other exhibits a Fano resonance \cite{Limonov2017} with a peculiar spectral dip beside the main peak. We can exclude such cases since none of them were seen in our entire experiment.

Figure 3(b) presents our least-square theoretical fitting for the emission spectra with $P(\omega)$. Because cavity 2 with $I_2=100 \ \upmu {\rm A}$ is considered nearly loss-compensated, we assume a low $\gamma_2$, setting it to 0.1 GHz to avoid any numerical problems like divergence. The data agree well with the experimental result, and the theoretical blue-side peak for $I_1 \gtrsim 5\ \upmu {\rm A}$ is slightly narrower mostly because of the neglected excitation of cavity 1 (see also Section 5 of Supplement 1). The analysis enables us to estimate the physical fitting parameters in the model, such as $\kappa$, $\gamma_1$ and $\delta$, which include the effect of the mode confinement factor. The eigenfrequencies ${\rm Re} \ \Delta \omega_i$ reconstructed with them, depicted by black points, ensure the correspondence between the sharp coalescence of the eigenmodes and the measured spectra.

Figure 3(c) and (d) show the $I_1$ dependence of estimated $\gamma_1$ and $\delta$. Here, the cavity coupling is found to be about $\kappa = 58 \ {\rm GHz}$ for the case of split resonances. Thus, $\kappa$ is fixed as that value in fitting the coalesced peaks for $I_1 \le 0.8 \ \upmu {\rm A}$, which are of more complexity (Section 7 of Supplement 1). The decline of $I_1$ monotonically enhances the material absorption in cavity 1 and hence $\gamma_1$. On the other hand, the reduction of the local carrier plasma effect \cite{Bennett1990} by decreasing $I_1$ induces a red shift there, which continuously diminishes $\delta$. Ideally, the EP should be near $\gamma_{1, {\rm EP}} = 2 \kappa \approx 116 \ {\rm GHz}$. Our measurement points have an interval of $\Delta I_1 = 0.2\ \upmu {\rm A}$ when $I_1$ is small, and $I_1 = 1.4\ \upmu {\rm A}$ is considered the closest to the EP. By carrying proper current $I_2 = 100 \ \upmu {\rm A}$ for cavity 2, we can cancel the detuning $\delta$ around the EP condition, which detrimentally lifts the degeneracy otherwise. Our device enables the efficient and fine control of its imaginary potential, with the thermal and carrier effects suppressed enough.

\section{LDOS enhancement by the EP degeneracy}
Since we have identified the fine condition of the EP degeneracy in our system, we are now able to examine its effects on the light emission. Figure 4(a) shows the measured spectral peak count as a function of $I_1$. Its single-bottomed property may look similar to the reversed pump dependence with the revival of lasing [Fig. 2(c)]. However, the peak intensity monotonically increases with $I_1$ declining below $2.4\ \upmu {\rm A}$, not below the estimated EP ($I_1=1.4\ \upmu {\rm A}$). This contradicts a na\"{i}ve speculation for the reversed pump dependence via the phase transition of ${\rm Im} \ \Delta\omega$ at the EP. When we look closely at the spectra, two peaks approach each other as $I_1$ decreases, and they are merged into a unimodal peak already at $I_1=2.4\ \upmu {\rm A}$, as displayed in blue in Fig. 4(c). If two non-mixing Lorentzian peaks were to be simply summed here like in a Hermitian system, the resultant contribution to the intensity must have saturated at the degeneracy with their peak frequencies coincident (Fig. S8 in Supplement 1). In addition, because the eigenstates for $I_1 > 1.4\ \upmu {\rm A}$ are supposed to be in the symmetric phase [$\gamma_1 < 2\kappa$ in Fig. 1 (b)], they become lossier, i.e. weaker when getting closer to the EP. Thus, the sharp growth of the peak count around the EP in Fig. 4(a), rather than the formation of its local minimum there, suggests the LDOS enhancement by the EP degeneracy that was predicted theoretically \cite{Lin2016,Pick2017gen,Pick2017enh} (Section 8 of Supplement 1).
\begin{figure}[t!]
	\centering\includegraphics[width=13.3cm]{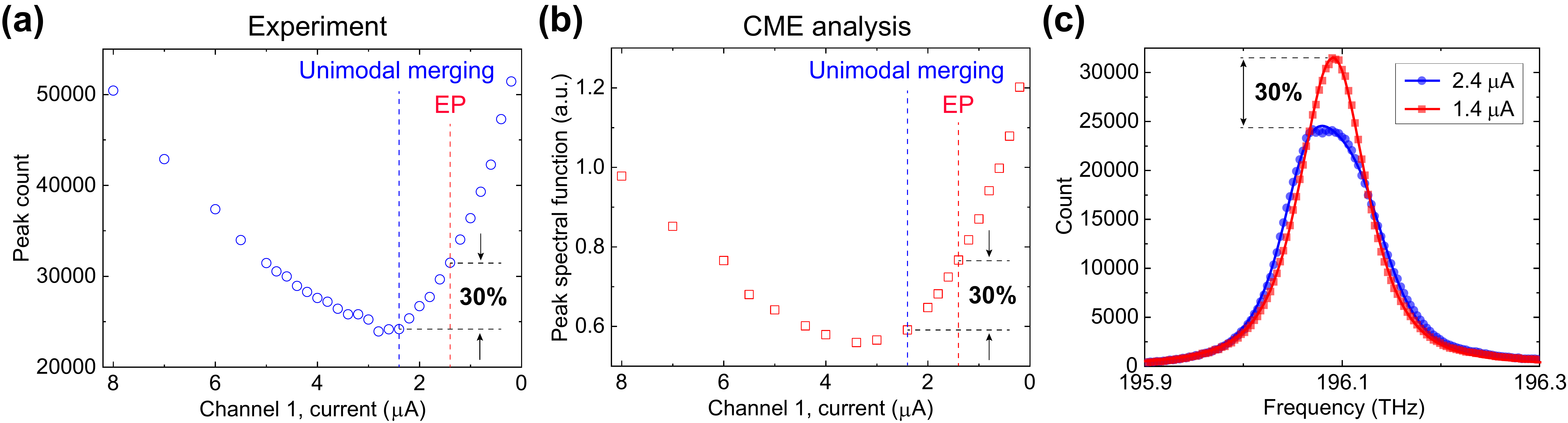}
	\caption{Transition of the spectral photon count. (a) Measured peak count of the device emission spectrum depending on $I_1$. (b) Peak of the theoretical spectral function $P \left( \omega \right)$ normalized with $\eta \gamma_{\rm cav} \left| c_2 \left( \omega \right) \right|^2 = 2\kappa / \pi$ and calculated with the estimated physical parameters for each $I_1$. (c) Experimental photon count spectra (symbols) and their CME fitting (solid curves) at the apparent unimodal merging of the two peaks ($I_1 = 2.4\ \upmu {\rm A}$, blue) and the near-EP condition ($I_1 = 1.4\ \upmu {\rm A}$, red). When $I_1$ is small, the excitation $c_1 (\omega)$ for cavity 1 is negligible. Thus, good consistency between $P(\omega)$ and the experimental data is obtained. Despite that the net loss of the eigenstates is intensified until the system's reaching the EP condition ($I_1 > 1.4\ \upmu {\rm A}$), the peak count grows sharply by 30\% from the two peaks' uniting ($I_1 = 2.4\ \upmu {\rm A}$) to the estimated EP ($I_1 = 1.4\ \upmu {\rm A}$). This indicates the LDOS enhancement based on the EP degeneracy.}
\end{figure}

To confirm the experimental anomaly of the peak count, we plot the peak of the normalized CME spectral function $P(\omega)$ including the obtained physical parameters for each $I_1$, in Fig. 4(b). Importantly, our analysis involves the interference of the two nonorthogonal spectral peaks mediated by non-Hermiticity, as we can see the equivalence between the CME spectral response and LDOS around the EP \cite{Pick2017enh} (Section 9 of Supplement 1). The theoretical peak intensity is consistent with the experimental data especially for $I_1 \le 3 \ \upmu {\rm A}$, where the excitation for cavity 1 is sufficiently small in experiment. Although we ensure the unimodal merging of the two coupled-mode peaks at $I_1=2.4\ \upmu {\rm A}$ [blue curve in Fig. 4(c)], the intensity here is close to its minimal value because of the enhanced loss, $\gamma_1 = 93.4 \ {\rm GHz} \approx 1.6 \kappa$. In contrast, it increases by 30\% until the near-EP condition with $I_1=1.4\ \upmu {\rm A}$ and a further larger $\gamma_1 = 115.6 \ {\rm GHz} \approx 2.0 \kappa$ in both theory and experiment, as the corresponding spectrum (colored red) is shown in Fig. 4(c). The monotonical increment in the peak intensity before the EP indicates the enhanced LDOS based on the EP degeneracy. Note that the enhancement ratio here is less than double seen in Fig. 1(d), because we control not $\kappa$ but $\gamma_1$ and the eigenstates for $I_1=2.4\ \upmu {\rm A}$ are already nonorthogonal.

Finally, the spontaneous emission spectra for $I_1=1.4\ \upmu {\rm A}$ and $0.2\ \upmu {\rm A}$ are fit by some distinct trial functions and plotted in both linear and semi-logarithmic scales as Fig. 5(a) and (b), respectively. Again, our CME spectral function reproduces the experimental data well, and the apparent discrepancy between them is just seen in the region with 10\% or less of the peak counts. The errors in their skirts can be mostly attributed to the slightly inclined background luminescence spectrum due to its peak located at around 1440 nm. This non-ideal factor can be corrected within the first order, as shown in Section 7 of Supplement 1. Remarkably, the entire section of the observed spectrum for $I_1=1.4\ \upmu {\rm A}$ is in accordance with the squared Lorentzian function abovementioned in Fig. 1(c) and (d), $4 \pi^{-1} C \left[ \gamma^2 / ({\Delta\omega}^2+\gamma^2)\right]^2$ with coefficient C, rather than with the ordinary Lorentzian function (see Section 6 of Supplement 1 for additional data). This evidences the resonance very near the exact EP and supports the enhancement of the photonic LDOS by the non-Hermitian degeneracy. We emphasize that the small difference between the CME analysis and squared Lorentzian response (LDOS) is rationalized by the fact that we measure not $\left| a_1(\omega) \right|^2$ but $\left| a_1(\omega) + i a_2(\omega) \right|^2$ (Section 1 and 8 of Supplement 1). Here, we can exclude the Voigt fitting function \cite{Stephan2005}, i.e. the convolution of the cavity Lorentzian factor and Gaussian noise, because it requires a too small average loss to have the EP ($26 \ {\rm GHz} < \gamma_{1, {\rm EP}}/2 = 58 \ {\rm GHz}$), as well as persistent Gaussian noise (27 GHz) inconsistently larger than our lasers' oscillation linewidths \cite{Takeda2013} (< 4 GHz: our finest measurement resolution; see Fig. S1 of Supplement 1). As $I_1$ further decreases down to $I_1=0.2\ \upmu {\rm A} \ (\gamma_1=\ 157.2 \ {\rm GHz} \approx 2.7 \kappa$), the experimental and best-fit CME spectra get settled in more Lorentzian shapes [Fig. 4(b), Section 9 of Supplement 1]. This indicates that the system loses the effect of the degeneracy on the LDOS for a large imaginary potential contrast, although the peak intensity further increases by the reduction of ${\rm Im} \ \Delta\omega$ and effective excitation of the dominant mode localizing at cavity 2.
\begin{figure}[t!]
	\centering\includegraphics[width=13cm]{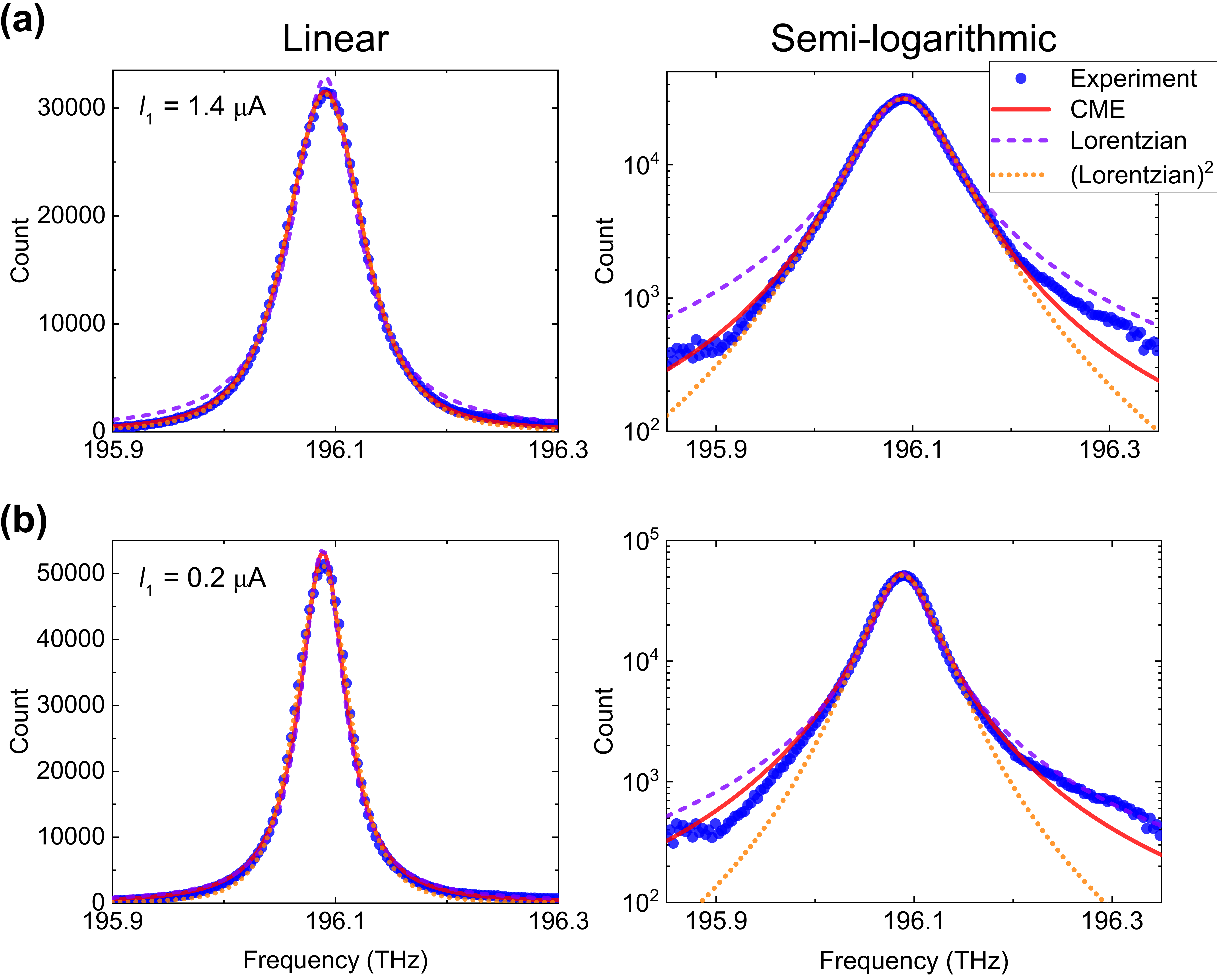}
	\caption{Device emission spectra near and far from the EP degeneracy. (a), (b), Observed photon count spectra (blue dots) and their theoretical fitting for (a) $I_1=1.4\ \upmu {\rm A}$ and (b) $I_1=0.2\ \upmu {\rm A}$. Left and right: their linear and semi-log plots, respectively. Our CME analysis (red line) explains both data well, and the plot in (a) agrees with a squared Lorentzian trial function (dotted orange curve) clearly better than a least-square Lorentzian trace (dashed purple curve), supporting the LDOS enhancement in the proximity of the EP. The emission with a smaller $I_1$ (b) comes to have a more Lorentzian component, as it is a state localizing at the heavily pumped cavity in the broken phase.}
\end{figure}

\section{Discussion and conclusion}
The EP resonance also exhibits a peculiar transient response. In our ideal EP condition, namely $\gamma_1 = 2 \kappa$, $\gamma_2 = 0$, $c_1 (\omega) = 0$, $|c_2 (\omega)|^2 = {\rm const.}$, cavity 1's radiation spectrum $\gamma_{\rm cav} |a_{\rm 1, EP}(\omega)|^2$ is a squared Lorentzian function [Eq. (S7) of Supplement 1]. Its inverse Fourier transform directly reflects the autocorrelation function, $C_{\rm 1, EP}(\tau) = \langle a_{\rm 1, EP}^{*}(t) a_{\rm 1, EP}(t+\tau) \rangle _t$. This measures the temporal average of field decay in cavity 1 during an interval of $\tau$, in response to every incoherent photon excited at cavity 2. In fact, the analytic operation yields $C_{\rm 1, EP}(\tau) \propto (1 + \kappa \tau)\exp (-\gamma_1 \tau /2)$, while the coupled modes in the Hermitian limit just undergo exponential loss, $C_{\kappa \gg \gamma}(\tau) \propto \exp (-\gamma_1 \tau /2)$.

Although the EP mode is distributed over both cavities, spontaneous emission occurs in cavity 2. As a result, it takes the time $1/\kappa$ for the fields to jump into cavity 1 and settle in the steady eigenstate. Here, the decay is prevented while the photons stay mostly in the loss-compensated cavity 2. Since $(1 + \kappa \tau) = (1 + \gamma_1 \tau /2) \approx \exp (\gamma_1 \tau /2)$ for $\gamma_1 \tau /2 \ll 1$, the net damping term $\exp (-\gamma_1 \tau /2)$ is indeed canceled within the first order of $\kappa \tau$. The EP hence enhances the peak spectral intensity, which corresponds to the integral of $C_{\rm 1, EP}(\tau)$. Note that this mechanism is also valid for the fields $a_{\rm 2, EP}$ of cavity 2. Thus, its radiation [Eq. (S9) of Supplement 1] and the entire device emission spectrum [Fig. 5(a), Fig. S6(a) of Supplement 1] hold the squared Lorentzian shapes. Exploiting such EP dynamics is an intriguing future direction.

Enhancing the peak LDOS at the passive EP will drastically modulate the photonic responses of quantum emitters \cite{Yang2020}, coherent absorbers \cite{Wong2016} and nonlinear optical devices \cite{Matsuda2011}. It can also have assistance of local gain \cite{Pick2017gen} and get further enhanced at a higher order EP \cite{Lin2016} with a ratio of $\sqrt{\pi} \Gamma(n+1) / \Gamma(n - 1/2)$ ($=4$ for $n=2$), where $n$ is its order and $\Gamma(n)$ is the gamma function. Nonlinear optical effects will even be made hundreds of times more efficient \cite{Pick2017enh} by adopting the non-Hermitian degenerate states. In addition, the reversed power dependence in the EP transition also shows nonlinearity on the pumping. This property provides us with new possibilities for nanophotonic switches and regulators.

Coupled nanolasers with electrical pumping, despite leading to the integration of periodic and controllable non-Hermitian optical systems, were not reported. Our buried heterostructure technology can provide such a framework and will open the door for access to singularities of group velocity \cite{Takata2017}, reconfigurable photonic topological insulators \cite{Takata2018}, and vortex charges and chirality of EPs \cite{Ota2020}. Large-scale passive devices \cite{Feng2013exp,Zhen2015,Kremer2019} in one and two dimensions successfully relax the condition of parameters for achieving EPs and rings of EPs. Nonetheless, fabrication-induced defects make it somewhat challenging to handle the degeneracy in such systems. Further study of corresponding active cavity arrays will also be of great significance.

In conclusion, we showed the clear EP transition of spontaneous emission with our current-injected photonic crystal nanolasers. We first clarified that it was difficult for lasing PT-symmetric eigenmodes to reach the EP degeneracy, because one of them was suppressed by the existence of cavity detuning. In contrast, the independent and efficient electrical pumping to our cavities enables the spontaneous emission near the exact EP, by limiting detrimental resonance shifts to the minimal level for active devices. In immediate proximity to the fine EP position elaborated by both our measurement and analysis, we found a squared Lorentzian emission spectrum, together with loss-induced growth of the peak power within the symmetric phase. These features demonstrate the peak LDOS enhancement that is intrinsic to the EP degeneracy. Our results represent an important step toward EP-based control of optoelectronic processes and large-scale non-Hermitian nanophotonic devices.
\section*{Funding}
Japan Science and Technology Agency (JST), CREST (JPMJCR15N4).

\section*{Acknowledgments}
We thank H. Sumikura and M. Takiguchi for their support with the measurement, H. Onji for CAD processing, and B. Zhen, A. Pick and J. W. Yoon for their kind discussions. This work was supported by Japan Science and Technology Agency (JST) through the CREST program under grant number JPMJCR15N4.

\section*{Disclosures}
The authors declare no conflicts of interest.




\section*{Supplementary material}

# Observing exceptional-point degeneracy of radiation with electrically pumped photonic crystal coupled-nanocavity lasers: supplemental document

## 1. Method details

### 1.1 Sample fabrication and design

The sample [Fig. 2(a) in the main text] contains an air-suspended InP photonic crystal slab and two InGaAlAs-based buried heterostructure nanocavities (red) with six quantum wells embedded. Here, an InAlAs sacrificial layer, the active layer and an overcladding InP layer were grown by metal-organic chemical vapor deposition (MOCVD). The heterostructures and airholes were patterned by electron-beam lithography with $SiO_2$ and $SiN$ mask layers, respectively. The periodic air holes and narrow trenches (black lines) were opened by inductively coupled plasma reactive etching (ICP-RIE). Selective wet chemical etching was carried out to define the nanocavities. After the regrowth of the intrinsic InP layer over the heterostructures, Si ion implantation followed by activation annealing and Zn thermal diffusion was applied to diagonally pattern n-doped and p-doped layers. Each of the resultant lateral PIN junctions is in contact with an InGaAs contact layer and Au-alloy metal pads at its edges. The diagonal doped layers and current blocking trenches help suppress leakage current between the electric channels.

Seven and five rows of airholes with a lattice constant of $a = 437\,\text{nm}$ are aligned on the outer sides of and between the line defects, respectively. A narrow line defect width of $0.85\sqrt{a}$ improves the modal confinement. The air hole radius and dimensions of the cavities are designed to be $R_\text{D} = 100\,\text{nm}$ and $2.185(5a) \times 0.3 \times 0.15\,\mu\text{m}^3$. The slab thickness is 250 nm.

### 1.2 Finite element simulation

The sample's eigenmodes were simulated with a commercial electromagnetic solver based on the finite element method (COMSOL Multiphysics). The fine structure of the buried nanocavities including the quantum wells were taken into consideration by using their effective index of $n_\text{BH} = 3.41$, and the refractive index of InP is $n_\text{InP} = 3.16$. No finite imaginary part of the index was assumed over the entire device. The computational domain was halved by a two-dimensional perfect magnetic conductor placed along the middle of the slab ($z = 0$). An air layer with a height of 3.5 μm was attached to the slab, and radiation loss was captured by the surrounding boundaries with the second-order scattering condition. We calculated the resonance frequencies and $Q$ factors of the coupled ground modes for different air-hole radii $R$, which can change depending fabrication conditions. The simulated wavelengths close to our experimental result were found for $R = 104.4\,\text{nm}$ (shown in the main text). The corresponding theoretical $Q$ factors are $2.5\times10^5$ and $2.7\times10^5$, while those of the first-order modes are less than 47,000. A similar device with a single nanocavity and the same parameters had the ground mode with a wavelength of 1530.8 nm and $Q = 2.9\times10^5$.

### 1.3 Measurement set-up

We use a caged measurement system with a device stage, a probe station, and a nano-motion lens revolver. The sample was placed on a device holder with a vacuum contact. The temperature of the holder was maintained at $25°\text{C}$ with its Peltier unit and a feedback controller.

Four electric microprobes were put on the metallic pads in contact with the ends of the doped layers. Independent DC currents for the right (channel 1) and left (channel 2) pairs of probes were applied and controlled with a precision source/measure unit. The device radiation was collected at the top with a 20X objective lens with a numerical aperture (NA) of 0.26 and coupled to an optical fiber. The near-field patterns [Fig. 2(b) and (e)] were observed with another 50X lens with $NA = 0.42$ and a near-infrared InGaAs camera.

In the I-L measurement [Fig. 2(c)], the voltage (and hence current) for one channel was swept upward and backward, with the injection current to the other kept constant. The output light passed through a variable filter with a bandwidth of 3 nm around 1530 nm and was measured with a low-noise power meter. Owing to vibrational and mechanical stability over the measurement system, we have not seen notable hysteresis behavior in the signal. Thus, we show the data without the points in the backward sweep for a better view.

The spontaneous emission (Fig. 3) was resolved with a spectrometer that has a grating with a groove number of 1000 g/mm and a spectral resolution of 0.12 nm. It was then detected by an InGaAs line detector array cooled down to −95°C with liquid nitrogen. The detector integration time was 60 s. The measurement was performed in the wavelength range of 1517 to 1542 nm, and the data from 195.5 to 196.7 THz were extracted for analyzing the ground-mode spectra. The minimum count in each curve, detected far away from the ground-mode resonances, was subtracted as the background component.

For the coherent radiation of the sample with the intense local pump (Fig. 4), we used a fiber-coupled optical spectrum analyzer. The wavelength sweep was performed under the lowest video bandwidth of 10 Hz for the maximum sensitivity, and every data point was averaged with 100 measurements. To compensate for negative detection currents (and power levels) caused by the spectral analyzer's small calibration error, the data were offset with reference to their minimum value in each sweep.

*1.4 Spectral coupled-mode analysis*

The Fourier transformation of Eq. (1) in the main text with the net external excitation terms

$$i\omega \begin{pmatrix} a_1(\omega) \\ a_2(\omega) \end{pmatrix} = \begin{pmatrix} i(\omega_0 + \delta) - \gamma_1 & i\kappa \\ i\kappa & i(\omega_0 - \delta) - \gamma_2 \end{pmatrix} \begin{pmatrix} a_1(\omega) \\ a_2(\omega) \end{pmatrix} + \begin{pmatrix} c_1(\omega) \\ c_2(\omega) \end{pmatrix}, \quad (S1)$$

is solved for $a_1(\omega)$ and $a_2(\omega)$ with $c_1 = 0$ to obtain Eq. (2). Considering our experimental conditions, the fitting function for the measured spontaneous emission spectra is

$$P(\omega) = \frac{(\gamma_1 + \kappa)^2 + (\Delta\omega - \delta)^2}{\left(\kappa^2 + \gamma_1\gamma_2 - \Delta\omega^2 + \delta^2\right)^2 + \left[(\gamma_1 + \gamma_2)\Delta\omega + (\gamma_1 - \gamma_2)\delta\right]^2} \eta\gamma_{cav} |c_2(\omega)|^2, \quad (S2)$$

where $\Delta\omega = \omega - \omega_0$, and $\eta\gamma_{cav}|c_2(\omega)|^2$ works as a proportional constant depending on the peak count. Here, $P(\omega)$ is affected by the radiation of both the excited and absorptive cavities. The resultant term $(\Delta\omega - \delta)^2$ in its numerator represents a small difference of spectral shapes between Eq. (S2) and the exact LDOS [1]. However, information of the LDOS is well reflected in the main parts of $P(\omega)$ and the observed spectra near the EP, where $(\Delta\omega - \delta)^2 < (\gamma_1 + \kappa)^2$.

In the fitting, we set $\gamma_2 = 0.1\,\text{GHz}$ because cavity 2 was considered to be transparent but not to provide gain. Note that the peak photon count started saturating, but the I-L data did not show any apparent reversed intensity (viz. oscillation) for $(I_1, I_2) = (0, 100\,\mu\text{A})$. Although we had to reduce the number of parameters due to the complexity of the problem, we were able to find good values of $\eta\gamma_{cav}|c_2(\omega)|^2$ ($\propto$ peak count) and $\omega_0$ (center of the peak structure). Thus, we performed the nonlinear least-square fitting for the rest of the variables, i.e. $\kappa$, $\delta$, and $\gamma_1$

with OriginPro. Here, the initial value of $\kappa$ was 60 GHz, considering the experimental and simulation result. Because $\kappa$, $\delta$, and $\gamma_1$ basically dominated the splitting, level difference and decaying tails of the spectral peaks, respectively, we were able to achieve a reliable estimation of them. Nonetheless, it became hard to have the consistent convergence of the fitting for the data after the peak coalescence. We hence fixed $\kappa$ as 58 GHz, as a value close to those obtained otherwise, and estimated $\gamma_1$ and $\delta$ for $I_1 < 1.0$ μA.

## 2. Single-laser sample

To estimate the optical properties of each diode in our two-cavity sample, we fabricated a device comprising a single nanocavity with the same structural parameter set, based on the previously reported technique [2]. Its false-color image is shown in Fig. S1(a). To well confine the ground mode of our mode-gap nanocavity, seven rows of triangular-lattice air holes are placed on both the upper and lower sides of the line defects, whose width is $0.85\sqrt{3}a$ ($a = 437$ nm). An active heterostructure designed to $2.185 \times 0.3 \times 0.15$ μm$^3$ (red) is buried at the center of the line defect. Trapezoidal p-doped (purple) and n-doped (green) layers adjacent to the cavity form a PIN junction, and they are in contact with the edges of $50 \times 50$ μm$^2$ metal pads at the top and bottom of the figure. An I-L curve of the device [Fig. S1(b)] indicates that its lasing threshold is $I_{th} \approx 37$ μA. Its emission spectra [Fig. S1(c), (d)] show that, as the injection current $I$ increases, an abrupt blue shift by the carrier plasma effect turns off around $I = I_{th}$, and then a linear red shift due to the thermal effect dominates. The consistent linewidth narrowing beyond the resolution (0.03 nm) of our optical spectral analyzer (OSA) together with the bright near-field patterns like ones seen in the two-laser experiment confirm that a high-$Q$ nanolaser is achieved. It is noteworthy that $Q = 14,000$ is obtained at $I = 37$ μA with the corresponding raw spectral data. The optical coherence of the emission from a similar heterostructure nanolaser was assured in Ref. 3. The threshold current here is larger than that seen in Ref. 2, because the doped layers are located closer to the heterostructure. Remarkably, their material absorptions help broaden the range of the loss (i.e. the on-site effective imaginary potential) of the nanocavity.

Figure S1(e) depicts the spectrum just above the threshold ($I = 42$ μA) and its fitting on a linear scale. The experimental data (orange points) contain non-negligible white spectral components and hence indicate that spontaneous emission is still effective, as well as weak stimulated emission. Thus, the data match up well with an offset Lorentzian function, which has a FWHM of 14.0 GHz (0.11 nm) and a background level of 0.183 pW (dashed purple curve). On the other hand, because its shape is asymmetric to the peak center, a Gaussian function $\propto \exp[-(\omega-\omega_C)^2/\sigma^2]$, based on $1/f$ flicker noise, can also fit the plot (dot-dashed curve in green). Importantly, the noise in this case comes from the fluctuation of active carrier population that is clamped around lasing [4]. Thus, the effect of the Gaussian noise on the line shape varies little and is *well below* the variance of the fitting in Fig. S1(e), $\sigma = 9.94$ GHz. We can hence neglect this factor for fitting Fig. 3 and 4 in the main text, where the loss-biased resonances, including the EP, have linewidths as large as several tens of gigahertz.

## 3. Current-in light-out curves

The I-L data for broader ranges of pumping under bidirectional sweeps are presented in Fig. S2. Because our measurement cage and equipment are connected with an optical fiber and a connector, mechanical and vibrational fluctuation in the set-up results in small discrepancies between the signal in the upward and downward sweeps. Nonetheless, we do not see any significant hysteresis behavior in the plots, showing that the mode competition and thermal

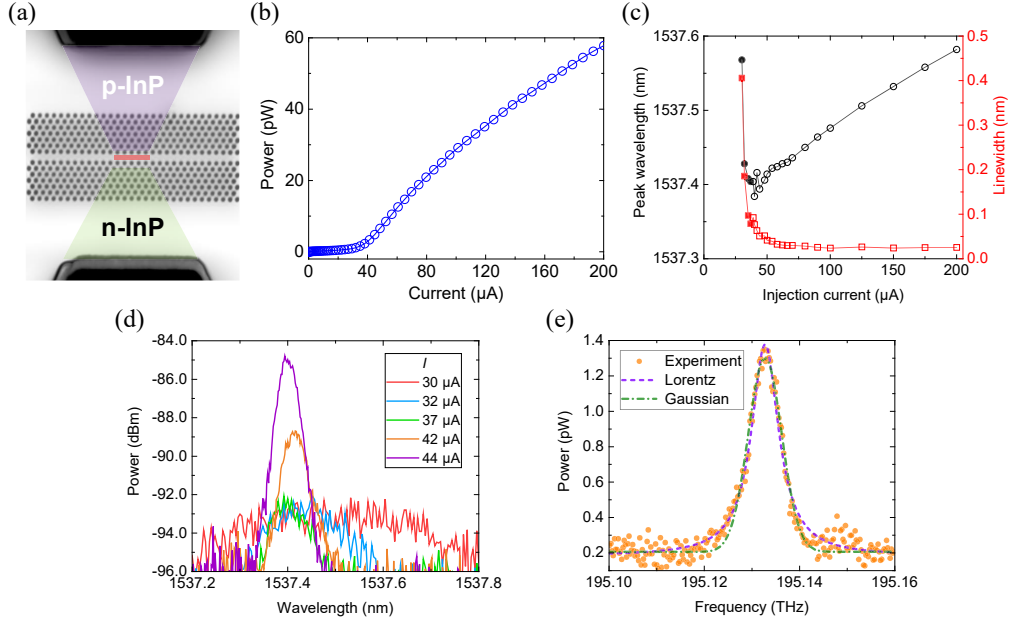

Fig. S1. (a) False-color image of the single-laser sample. (b) I-L curve of the device. The oscillation threshold is $I_{th} \approx 37$ μA. (c) Spectral peak wavelength (black) and linewidth (red) of the emission, dependent on the injection current $I$. Filled markers: estimations with moving average of the spectra. (d) Device emission spectra around $I_{th}$. The data for $I = 37$ μA (green), without significant amplification by oscillation, indicate $Q = 14,000$. (e) The linearly scaled spectrum for $I = 42$ μA with offset Lorentzian and Gaussian fitting.

nonlinearity are well suppressed in the sample. When $I_2$ is fixed and $I_1$ is swept [Fig. S2(a), also shown in the main text], we systematically see the steep suppression and revival of the coherent radiation power in diminishing $I_1$ for $I_2 \geq 300$ μA. In comparison, the rise in power for the case of constant $I_1$ after oscillation is gentle. Fig. S2(a) also indicates that the system with $I_2 = 100$ μA and $I_1 \leq 10$ μA is in the spontaneous emission regime. On the other hand, the reversed pump dependence for the other case [Fig. S2(b)] is less prominent. This is probably because the cavities are detuned more in this condition; for example, we have a solitary resonance of cavity 1 at 1529.47 nm for $(I_1, I_2) = (300\ \mu\text{A}, 0)$. This mode is off from the coalescence at 1529.9 nm under the constant $I_2 = 100$ μA (Fig. 3).

## 4. Symmetrically pumped two-laser sample

To estimate the cavity coupling $\kappa$, we also analyze the emission spectrum of the symmetrically pumped system with $I_1 = I_2 = 30$ μA, which is below the oscillation threshold (Fig. S3). The experimental data (blue dots) show the two coupled-mode peaks with $Q$ factors of about 4,000 based on their FWHM ($\approx 0.37$ nm). Because both channels have the same injection current, we need to take into consideration the net field excitation $c_1(\omega)$ for cavity 1 that is neglected in the main text. The solution of the coupled mode theory (CMT) is hence modified as,

$$a_1(\omega) = \frac{\left[\gamma_2 + i(\Delta\omega + \delta)\right]c_1(\omega) + i\kappa c_2(\omega)}{\kappa^2 + \left[\gamma_1 + i(\Delta\omega - \delta)\right]\left[\gamma_2 + i(\Delta\omega + \delta)\right]}, \quad (S3)$$

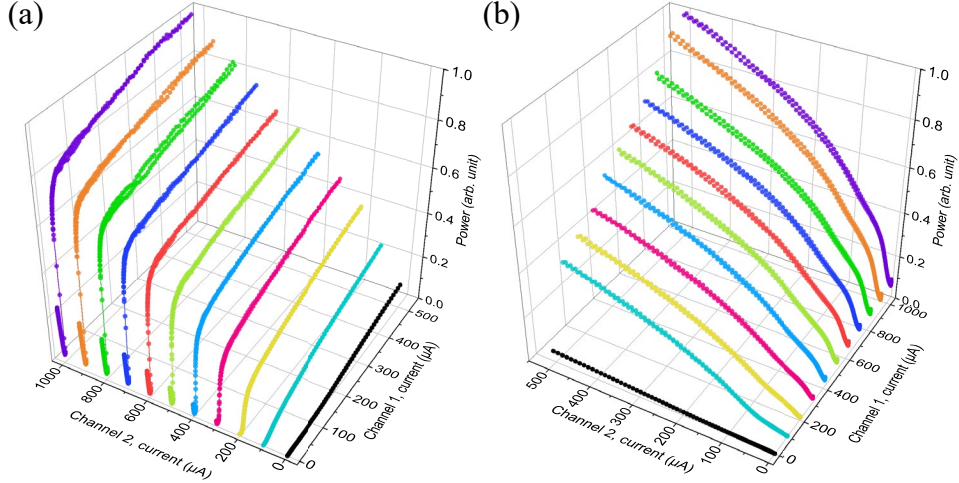

Fig. S2. (a) I-L plots of the sample for swept $I_1$ and fixed $I_2$ and (b) those for varied $I_1$ and constant $I_2$. The signal is filtered around 1530 nm with a bandwidth of 3 nm. The integration time of the power meter is 1 s.

$$a_2(\omega) = \frac{i\kappa c_1(\omega) + [\gamma_1 + i(\Delta\omega - \delta)]c_2(\omega)}{\kappa^2 + [\gamma_1 + i(\Delta\omega - \delta)][\gamma_2 + i(\Delta\omega + \delta)]}, \quad (S4)$$

and we again consider $P(\omega) = \eta\gamma_{cav}|a_1(\omega) + ia_2(\omega)|^2$ as our trial function.

If we assume that the net pumping is identical for the two cavities, we reasonably have $\gamma_1 \approx \gamma_2$ and $|c_1(\omega)| \approx |c_2(\omega)|$. In that case, however, finite $\delta$ does not induce any level margins of the two peaks, despite that the experimental result exhibits a 20% discrepancy between the peak counts. This means that the two cavities have imbalanced pumping possibly due to inhomogeneous electric contact among the four DC probes and metal pads and/or due to a difference in current leak paths of the channels. A best-fit theoretical curve is obtained under a condition that the ratio of the field excitation rates is inversely proportional to that of the on-site field loss rates, namely

$$r \equiv \frac{|c_2(\omega)|}{|c_1(\omega)|} = \frac{\gamma_1}{\gamma_2}, \quad (S5)$$

which is a rational indication of the linear loss model.

Another important factor is that, $c_1(\omega)$ and $c_2(\omega)$ are not expected to have any phase coherence, since they are spontaneous emissions from distinct quantum wells and cavities (in sharp contrast to the interference between the two cavity sources $a_1$ and $a_2$ under the single point excitation). Thus, we take the unweighted average of $P(\omega)$ for their phase differences spanning over $[0, \pi)$ at even intervals. We eventually find theoretical data that agree with the experiment in the peak counts, the ridge between the peaks, and their skirts outside (Fig. S3), which also suggests a good estimation of the parameters, $\kappa = 61$ GHz, $\delta = 22$ GHz, $\gamma_2 = 18.3$ GHz and $r = 0.82$.

It is notable that a small change in the coupling $\Delta\kappa$ up to a few gigahertz, together with a compensation of the detuning by $\Delta\delta = -2\Delta\kappa$, little affects the theoretical curve. We also see broader linewidths of the experimental peaks than those of the theory. This indicates small extra Gaussian noise, which gets negligible in the case of asymmetric pumping and high local loss.

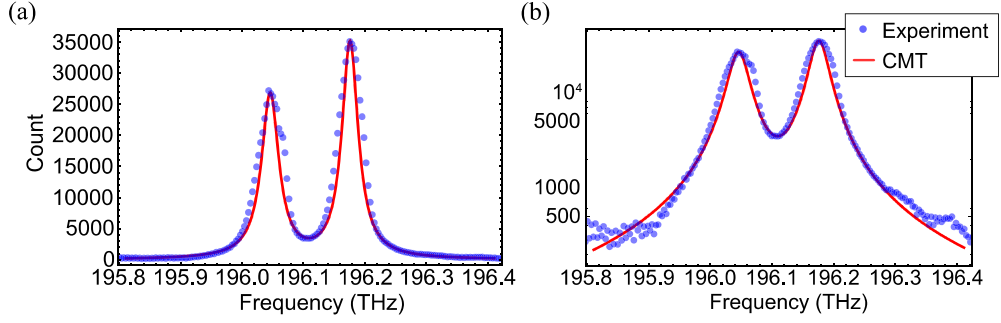

Fig. S3. The spectral photon counts for $I_1 = I_2 = 30$ μA in (a) linear and (b) semi-logarithmic plots. Blue dots: experimental data. Red curve: best CMT fitting with Eqs. (S3)-(S5) for $\kappa = 61$ GHz, $\delta = 22$ GHz, $\gamma_2 = 18.3$ GHz and $r = 0.82$. The discrepancy between the experiment and analysis is attributed to small Gaussian random noise, which becomes minor in the spontaneous emission measurement with the asymmetric pumping and high local loss factors (Fig. 3 and 5).

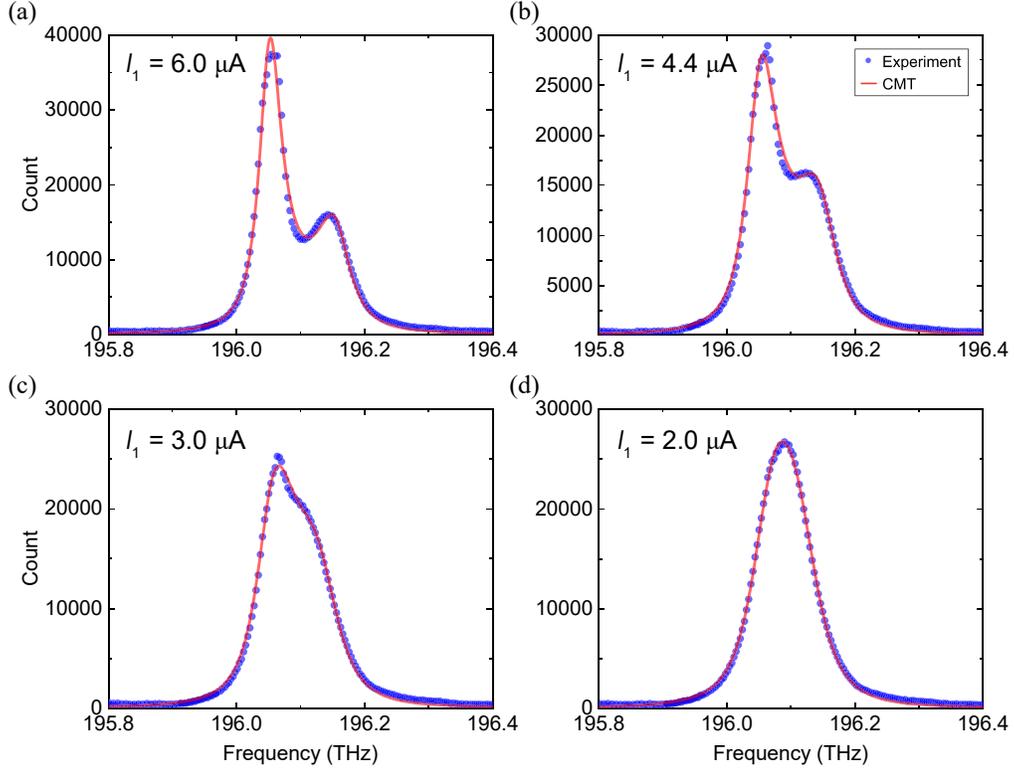

Fig. S4. Spontaneous emission spectra of the sample (blue dots) and their CMT fitting (red lines). (a) $I_1 = 6.0$ μA, (b) $I_1 = 4.4$ μA, (c) $I_1 = 3.0$ μA and (d) $I_1 = 2.0$ μA. $I_2 = 100$ μA.

## 5. Measured spectra in spontaneous emission regime

In Fig. S4, we present experimental photon count spectra for some different $I_1$ and $I_2 = 100$ μA (blue dots) with their CME fitting (red curves), to show that our model consistently reproduces the measured data well. Here, there is small difference between them,

in terms of the ridge lines and lower peaks, because we ignore $c_1(\omega)$. However, the fitting including it leads to significant complexity of the task, since the injection current to the channels does not necessarily reflect the ratio of the excitation amplitudes, as inferred in Sec. III. Nonetheless, the flat-top [Fig. S4(a)] and pointed [Fig. S4(b), (c)] peak structures in the experimental data can simply be attributed to the fluctuation of the detection.

The emission spectra around the EP condition, together with the fitting using different trial functions, are depicted in Fig. S5. Because the transition of the system spectral response is continuous, we can also observe its particular EP-based property nearby. For both $I_1 = 1.6\,\mu\text{A}$ [Fig. S5(a)] and $I_1 = 1.2\,\mu\text{A}$ [Fig. S5(b)], the Lorentzian curves obviously diverge from the experimental data in peak counts and decaying tails. In contrast, our CME and squared Lorentzian functions explain them significantly better and hence demonstrate the enhanced LDOS. In our measurement condition where both $a_1(\omega)$ and $a_2(\omega)$ are collected, the radiation at the EP has slightly broader spectral skirts than the squared Lorentzian function [see Eq. (S2) and Methods]. Thus, Fig. S5(a) signifies the system ($\gamma_1$) below the EP. The narrower peak shown in Fig. S5(b) is slightly more compliant with the Lorentzian fitting compared with that for $I_1 = 1.4\,\mu\text{A}$ (Fig. 5), suggesting that the system with $I_1 = 1.2\,\mu\text{A}$ ($\gamma_1 = 119.7\,\text{GHz}$) is just beyond the EP.

## 6. Correction of background counts in the spontaneous emission spectra

The background emission spectrum is not completely flat, because the electroluminescence of the sample has a global spectral peak located around 1440 nm. We can correct this non-essential factor within the first order in the following way, for clearer comparison between the theoretical and experimental results.

We first pick a raw spectral curve from 195.0 to 197.0 THz and take its moving average for denoising. Next, we draw a smooth lower envelope of the resultant data, which avoids the effect of the resonant peaks. Linear fitting of the envelope gives the background gradient that should be compensated. The linear function with this slope and reference to the least point is subtracted from the entire raw data, before the additional offsetting with the minimum count.

Figure S6 shows the experimental spontaneous emission spectra with the linear background correction for (a) $I_1 = 1.4\,\mu\text{A}$ and (b) $I_1 = 0.2\,\mu\text{A}$ and their theoretical fitting. Compared with Fig. 5, the corrected data here agree better with our CMT fitting in both (a) and (b), especially on the right skirts of the peaks. The discrepancy between them is just within the order of hundreds of counts, and it is mostly attributed to the small oscillations superposed over the data, which have an amplitude of about 100 counts and correspond to the Fabry-Perot resonances by the line defects. We emphasize that the difference between the CMT and squared Lorentzian function around the EP condition [Fig. S6(a)] comes from the radiation from the excited cavity 2 that is subject to the reaction of cavity 1 and hence broadens the spectral tails via the quadratic term $(\Delta\omega - \delta)^2$ in the response function. It is also noteworthy that the CMT also explains the data for $I_1 = 0.2\,\mu\text{A}$ best of the considered trial functions [Fig. S6(b)], and it becomes closer to the Lorentzian function as the system gets away from the EP.

## 7. Estimated parameters

We discuss the rest of the parameters estimated in our theoretical analysis and shown in Fig. S7. The characteristics of the cavity-mode wavelengths $\lambda_1(I_1)$ and $\lambda_2(I_1)$ [Fig. S7(a)] include the information about the average frequency $\omega_0$ and cavity detuning $\delta$. While the resonance of cavity 2 under the constant pumping ($I_2 = 100\,\mu\text{A}$) remains near $\lambda_2 \approx 1529.9$ nm in the symmetric phase, the other mode ($\lambda_1$) undergoes a continuous red shift with decreasing $I_1$ due

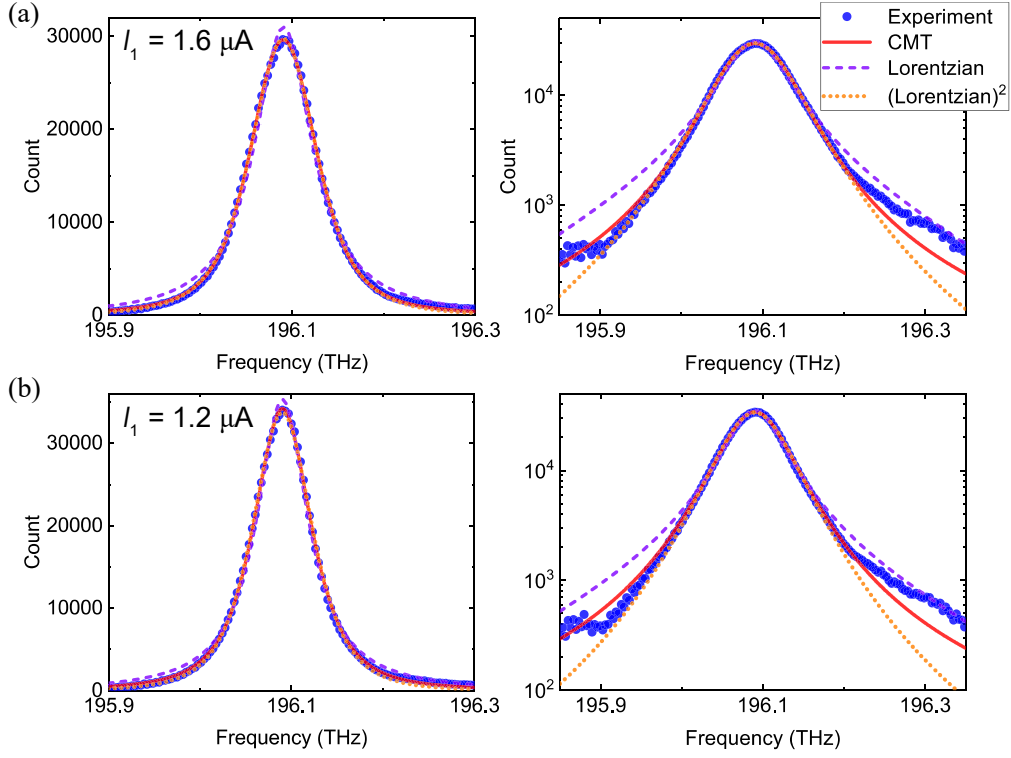

Fig. S5. Measured spontaneous emission spectra near the EP condition and corresponding theoretical fitting. (a) $I_1 = 1.6\,\mu\text{A}$ and (b) $I_1 = 1.2\,\mu\text{A}$. Left and Right: the data in linear and semi-logarithmic plots, respectively.

to the suppression of the carrier plasma effect. The detuning almost vanishes around $I_1 = 1.4\,\mu\text{A}$, i.e. the EP condition, and its sign is flipped beyond the degeneracy. The small but consistent red shift of $\lambda_2$ for $I_1 \leq 1.6\,\mu\text{A}$ is present possibly because of heat caused by mode localization or weak carrier depletion due to amplified spontaneous emission. Note that the absorption loss ($Q$ factor) of the major localized eigenmode falls (scales) rapidly in the broken phase. The values of the coupling rate $\kappa$ are mostly distributed between 58 and 59 GHz over our analysis [Fig. S7(b)]. As the fitting for the coalesced peaks with narrowing linewidths becomes difficult, we approximate the coupling as $\kappa = 58$ GHz considering the average, and analyze the data by estimating $\gamma_1$ and $\delta$ for $I_1 \leq 0.8\,\mu\text{A}$.

## 8. Contradiction of diabolic point picture with peak enhancement around EP

We show that the Hermitian diabolic point picture cannot explain the monotonical increase in the spectral peak power around the EP degeneracy, which has been observed in our experiment. If the controlled electric pumping would not induce any non-Hermitian features in our system, the eigenmodes would be orthogonal to each other and have Lorentzian spectral shapes. Thus, the total radiation spectrum would be proportional to the sum of two Lorentzian functions. Here, we consider an example with a common net decay rate $\gamma = 47$ GHz and normalization factor $A = \gamma$, which depends on the peak splitting (frequency difference) $\Delta$,

$$P(\omega, \Delta) = A\left(\frac{\gamma}{(\omega - \Delta)^2 + \gamma^2} + \frac{\gamma}{(\omega + \Delta)^2 + \gamma^2}\right). \quad (S6)$$

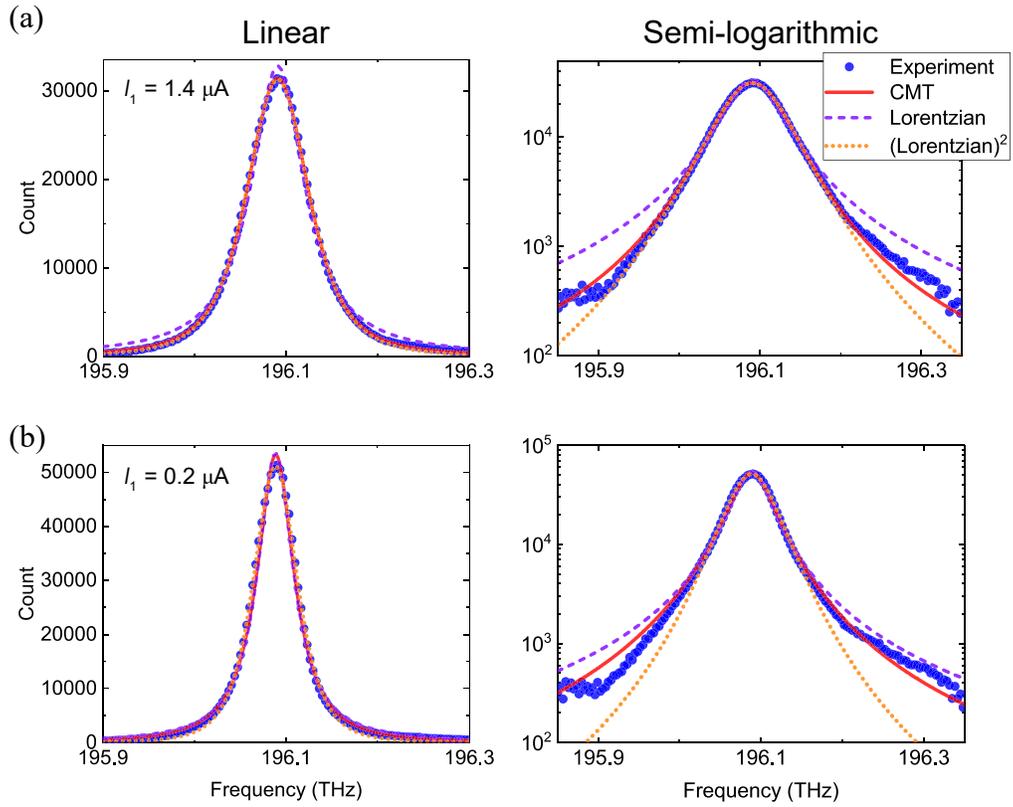

Fig. S6. Device spontaneous emission spectra with the linear background correction, to be compared with Fig. 5 in the main text. (a) $I_1 = 1.4\,\mu\text{A}$ and (b) $I_1 = 0.2\,\mu\text{A}$. Linear functions with slopes of 240 and 200 counts/THz are subtracted from the raw data for (a) and (b), respectively.

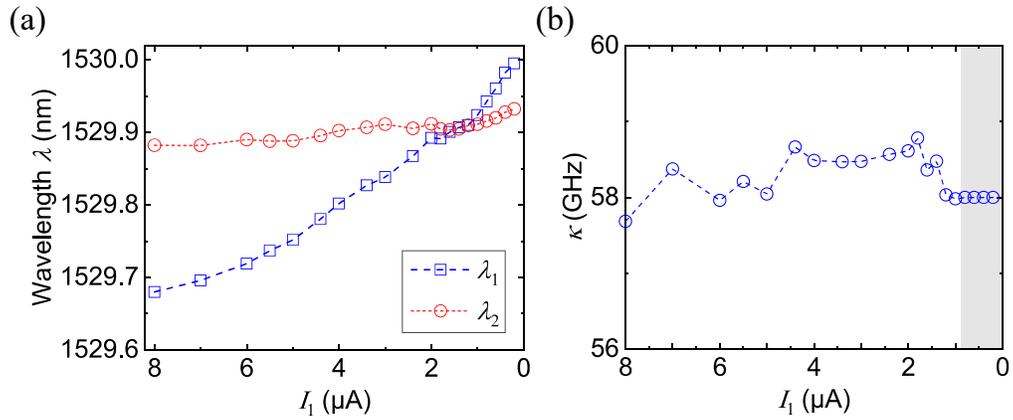

Fig. S7. Estimated dependence of the (a) resonance wavelengths of the cavity modes $(\lambda_1, \lambda_2)$ and (b) cavity coupling rate $(\kappa)$ on $I_1$ in the spontaneous emission experiment. In the shaded area of (b), $\kappa$ is approximated as 58 GHz to obtain a reliable fitting result.

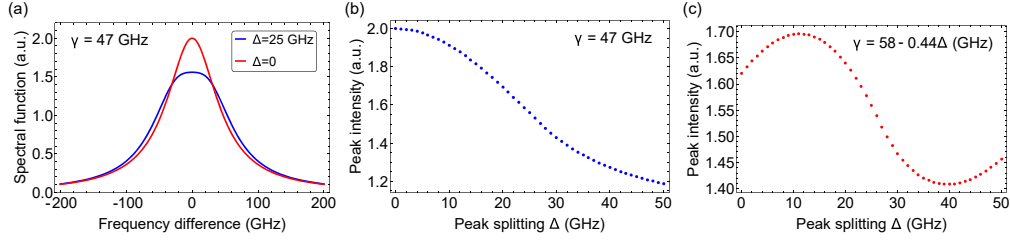

Fig. S8. Accidental degeneracy (Hermitian diabolic point) of coupled modes with Lorentzian spectra. (a) Sum of two orthogonal Lorentzian resonances $P(\omega, \Delta)$ with their common net loss $\gamma = 47\,\text{GHz}$, for different peak splitting $\Delta$ ($\Delta = 0$: diabolic point). (b), (c) Peak intensity of $P(\omega, \Delta)$ dependent on $\Delta$, for (b) constant $\gamma = 47\,\text{GHz}$ and (c) variable $\gamma = 58 - 0.44\Delta\,\text{GHz}$. Both (b) and (c) are not consistent with the experimental peak enhancement around the EP.

Here, $A$ is constant because the major excitation for the system is fixed, $I_2 = 100\,\mu\text{A}$. The value of $\gamma$ comes from $(\gamma_1 + \gamma_2)/2 \approx \gamma_1/2$ for $I_1 = 2.4\,\mu\text{A}$ in our experiment, where the observed spectrum exhibits an apparent unimodal coalescence (c.f. Fig. 4(c) in the main text).

As seen in Fig. S8(a), a finite $\Delta = 25$ GHz enables the formation of a unimodal peak based on the Lorentzian functions, and the peak intensity rises until the accidental degeneracy of frequency by the orthogonal modes, i.e. $\Delta = 0$ (Hermitian diabolic point). However, the intensity saturates when $\Delta$ approaches zero, as shown in Fig. S8(b). This is because, differentials of Lorentzian functions always vanish at their peaks ($\omega = +\Delta$ and $\omega = -\Delta$ in the first and second terms of Eq. (S6)), and these spectral convergences become exactly coincident at $\omega = 0$ when $\Delta = 0$. Note that such saturation occurs for any summation of spectral functions that are derived by Fourier (integral) transform of continuous dynamics. The monotonical increase in the peak power around the EP, clarified in our work (Fig. 4(a) and (b)), cannot be explained without the LDOS enhancement based on the EP degeneracy of eigenstates.

Furthermore, $\gamma_1$ actually becomes larger from the unimodal peak formation ($\gamma_1/2 \approx 47$ GHz) to complete degeneracy ($\gamma_1/2 \approx 58$ GHz), because we decrease $I_1$ in experiment. If we take this into consideration with a negative correlation between $\Delta$ and $\gamma$ covering $(\Delta, \gamma) = (0, 58\,\text{GHz})$ and $(25\,\text{GHz}, 47\,\text{GHz})$, $P(\omega, \Delta)$ exhibits a minimum at $\Delta = 0$, as shown in Fig. S8(c). This comes from the peak power reduction by linewidth broadening and indicates a more serious contradiction of the diabolic point picture with our experimental result.

## 9. Some theoretical properties of the system's spectral response

The photonic LDOS is directly relevant to how much a point-source dipole excitation couples with possible harmonic modes in the system [5]. In the spectral coupled-mode model here, the field solutions $\{a_1(\omega), a_2(\omega)\}$ denote the response of the coupled eigenmodes to the net excitations $\{c_1(\omega), c_2(\omega)\}$ for the tiny cavities. Because the cavity coupling $\kappa$ is approximated as a constant for the considered frequency range, it is reasonable that the intensity of cavity 1 $|a_1(\omega)|^2$ under the excitation only for transparent cavity 2 $[c_1(\omega) = 0, \gamma_2 = 0]$ exactly involves the peculiar squared Lorentzian LDOS [1] at the EP condition $(\delta = 0, \kappa = \gamma_1/2)$,

$$\left|a_1(\omega)\right|^2_{\text{EP}} = \left(\frac{\gamma_1}{2}\right)^2 \frac{\left|c_2(\omega)\right|^2}{\left[\left(\frac{\gamma_1}{2}\right)^2 + \Delta\omega^2\right]^2}, \qquad (S7)$$

and vice versa for $c_2(\omega) = 0, \gamma_1 = 0$, $\delta = 0$ and $\kappa = \gamma_2 / 2$,

$$\left|a_2(\omega)\right|^2_{\text{EP}, c_2=0} = \left(\frac{\gamma_2}{2}\right)^2 \frac{\left|c_1(\omega)\right|^2}{\left[\left(\frac{\gamma_2}{2}\right)^2 + \Delta\omega^2\right]^2}, \quad (S8)$$

as mentioned in Ref. 6. Remarkably, the full width at half maximum (FWHM) of Eq. (S7) [(S8)] is $\sqrt{\sqrt{2}-1}\gamma_1(\gamma_2) \approx 0.644\gamma_1(\gamma_2)$, where $\gamma_1(\gamma_2)$ is that for the Lorentzian function with the same loss factor. This shows the *linewidth narrowing* of the EP resonance.

Without loss of generality, we focus on the former case as we have done in the experiment. Here, the excited cavity (#2) has an intensity additionally affected by $\Delta\omega^2$, as the back action from the cavity 1's resonance,

$$\left|a_2(\omega)\right|^2_{\text{EP}} = \frac{\left(\gamma_1^2 + \Delta\omega^2\right)\left|c_2(\omega)\right|^2}{\left[\left(\frac{\gamma_1}{2}\right)^2 + \Delta\omega^2\right]^2}. \quad (S9)$$

Around the EP frequency ($\omega \approx \omega_0$ or $\Delta\omega^2 \ll \gamma_1^2$), however, Eq. (S9) also features a squared Lorentzian shape, holding consistency with the result of the singular perturbation analysis [1]. This is also the case for our fitting function, $P(\omega) = \eta\gamma_{\text{cav}}\left|a_1(\omega) + ia_2(\omega)\right|^2$.

The device emission in the large loss limit ($\gamma_1 \gg \kappa$) is another important property. Cavity 1's intensity, with $c_1(\omega) = 0, \gamma_2 = 0$, and $\delta = 0$ for simplicity, can reduce to

$$\left|a_1(\omega)\right|^2 = \frac{\kappa^2}{\gamma_1^2} \frac{\left|c_2(\omega)\right|^2}{\Delta\omega^2 + \frac{\kappa^4}{\gamma_1^2}\left(1 - \frac{\Delta\omega^2}{\kappa^2}\right)^2} \approx \frac{\kappa^2}{\gamma_1^2} \frac{\left|c_2(\omega)\right|^2}{\Delta\omega^2 + \left(\frac{\kappa^2}{\gamma_1}\right)^2}, \quad (S10)$$

when the main peak structure ($\left|\Delta\omega\right| \leq \kappa^2/\gamma_1$) is sufficiently narrower than the width of the quadratic factor $[1-(\Delta\omega^2/\kappa^2)]$, i.e. $\gamma_1 \gg \kappa$. Because the same derivation is applicable for $P(\omega)$, a Lorentzian emission spectrum with a FWHM of $2\kappa^2/\gamma_1$ is expected in the case of a high imaginary potential contrast.

We also point out that the position of the emission peaks is not identical to the eigenfrequencies of the effective Hamiltonian. Approximate peak frequencies of $P(\omega)$, are those giving extremal values of its denominator and found as real solutions for the following cubic equation,

$$2\Delta\omega^3 + \left(\gamma_1^2 - 2\kappa^2 - 2\delta^2\right)\Delta\omega^2 + \gamma_1^2\delta = 0. \quad (S11)$$

They successfully take into consideration the effect of the overlap between the two spectral peaks with their level difference and finite linewidths, and provide the exact result for $\left|a_1(\omega)\right|^2$ under a flat (constant) excitation spectrum. We have numerically assured that Eq. (S11) and the estimated parameters reproduce the peak position of the measured spectra well. In contrast, the eigen-wavelengths of the lower peaks in Fig. 3(b) somewhat diverge from the peak locations of the experimental data, because of the skirts of the other peaks with larger peak counts.

A simpler example is the case where $\delta = 0$. Here, we obtain the analytic peak solutions $\Delta\omega = \pm\sqrt{\kappa^2 - (\gamma_1^2/2)}$, which are certainly different from the corresponding eigenfrequencies, $\Delta\omega_i = \pm\sqrt{\kappa^2 - (\gamma_1^2/4)}$.

## 10. Eigenvalue analysis for lasing states

To analyze the eigenfrequencies of the lasing states, the time-domain CMEs [Eq. (1) in the main text] are solved under the condition that cavity 1 has a loss $\gamma_1$ and cavity 2 provides a gain for oscillation, $\gamma_2 = -g_2$. Here, while $\gamma_1$ depends on $I_1$, $g_2$ is adaptively determined so that the net gain and loss of each eigenstate are cancelled, when $I_2$ is large enough. If we assume that all the parameters and the eigenfrequencies $\omega_e$ are real (i.e. the eigenmodes are lasing), the characteristic equation for $\omega_e$ can be divided into its real and imaginary parts as,

$$\text{Re}: (\omega_e - \omega_1)(\omega_e - \omega_2) + \gamma_1 g_2 - \kappa^2 = 0, \quad \text{(S12)}$$

$$\text{Im}: (\omega_e - \omega_1)g_2 - (\omega_e - \omega_2)\gamma_1 = 0, \quad \text{(S13)}$$

where $\omega_1 = \omega_0 + \delta$ and $\omega_2 = \omega_0 - \delta$. Here, the latter equation gives the clamped oscillation-threshold gain that is supplied by cavity 2,

$$g_2 = \frac{\omega_e - \omega_2}{\omega_e - \omega_1}\gamma_1. \quad \text{(S14)}$$

This enables us to eliminate $g_2$ in the former equation [7]. The resultant cubic equation for $\omega_e$ is,

$$(\omega_e - \omega_1)^2(\omega_e - \omega_2) + \gamma_1^2(\omega_e - \omega_2) - \kappa^2(\omega_e - \omega_1) = 0. \quad \text{(S15)}$$

Our analysis for Fig. 2(d) in the main text takes into consideration the following parameter-dependent properties. First, we assume the saturable loss $\gamma_1(I_1)$ in cavity 1 as a two-step linear model comprising straight lines connecting $(I_1, \gamma_1) = (-5\,\mu\text{A}, 160\,\text{GHz})$, $(8\,\mu\text{A}, 33\,\text{GHz})$ and $(50\,\mu\text{A}, 0.2\,\text{GHz})$, which is close to that estimated by our spontaneous emission measurement and fitting [Fig. 3(c)].

Second, we isolate the red shift $\tau_T(I_1)$ caused by the sample's thermal expansion. This effect can be safely regarded as the proportional variation of all the eigenfrequencies on the pumping current [2], and the contribution of $I_1$ is given by,

$$\tau_T(I_1) = C_T I_1 = -8.547 \times 10^{-2} I_1 \text{ (GHz)}. \quad \text{(S16)}$$

Here, $I_1$ is in microamperes and the coefficient $C_T [\text{GHz}/\mu\text{A}]$ is calculated with the data points with the largest $I_1$, where $\gamma_1$ is well saturated.

Finally, the reduction of the carrier plasma effect $\tau_{C,L}(I_1)$ for the restored lasing state in cavity 2 is approximated as a proportional dependence on the peak power $n_P(I_1)$ in pW [Fig. 5(b)], since $I_2$ is constant and the steady active carriers in cavity 2 are consumed one by one by the stimulated photons localizing there. This is only applied for $|\lambda_-\rangle$ with $I_1 \leq 5.4\,\mu\text{A}$ and shown as

$$\tau_{C,L}(I_1 : I_1 \leq 5.4\,\mu\text{A}) = C_C n_P(I_1) = -2.726 \times 10^{-1} n_P(I_1) \text{ (GHz)}, \quad \text{(S17)}$$

where $C_C [\text{GHz}/\text{pW}]$ is estimated with the lasing frequencies with minimum $I_1$, where the eigenstates are deeply in the broken phase.

With all of them, we eventually reach our trial function,

$$\omega(I_1) = \omega_e\left[\gamma_1(I_1)\right] + \tau_T(I_1) + \tau_{C,L}\ (I_1 : I_1 \leq 5.4\,\mu A), \qquad (S18)$$

and we use $\kappa = 71\,\text{GHz}$, $\omega_1 = 196.1143\,\text{THz}$ (1529.72 nm) and $\omega_2 = 196.1284\,\text{THz}$ (1529.61 nm) in our fitting. A somewhat large value of $\kappa$ is adopted here, since we have to average the effect of the cavity detuning, which is hard to be precisely identified. Because the equation for $\omega_e \in \mathbb{R}$ is cubic, the solution can include the eigenstates in both the exact and broken phases. The analytic result is shown as the black dots in Fig. 2(d) and explains a major portion of the experimental data.

In Fig. S9, we compare the transition of the theoretical eigen-wavelengths (measured in air) and threshold gain in the lasing regime for the case of no and finite cavity resonance detuning $\delta$. Here, we exclude the change of heat and carrier contributions to the resonances in the process and just measure the impact of gain, loss and constant (average) detuning, by solving the algebraic equation (S15) for the eigenfrequencies $\omega_e \in \mathbb{R}$. We consider their dependence on $\gamma_1$, without converting it into the injection current $I_1$. The cavity coupling here is $\kappa = 71\,\text{GHz}$, as adopted for fitting Fig. 2(d). Note that the adaptive threshold gain $g_2$ is given by Eq. (S14) for each $\omega_e$. Complex solutions ($\text{Im}\,\omega_e \neq 0$) are ruled out in the analysis, because they indicate unstable decaying states.

When the local cavity-mode wavelengths are identical, $\lambda_1 = \lambda_2 = 1529.665\,\text{nm}$, the system holds the singular EP degeneracy of the two dispersive solutions $\{|\lambda_+\rangle, |\lambda_-\rangle\}$ corresponding to the coupled modes, even with the variable gain due to heavy pumping [Fig. S9(a)]. The red-side ($|\lambda_+\rangle$) and blue-side ($|\lambda_-\rangle$) branches in the symmetric phase have the same gain $g_2 = \gamma_1$, until they reach the EP at $\gamma_{1,\text{EP}} = \kappa = 71\,\text{GHz}$ [Fig. S9(c)]. Thus, both states are ideally stable, and the device hence does not exhibit selective excitation for either $|\lambda_+\rangle$ or $|\lambda_-\rangle$. On the other hand, the additional solution $|\lambda_B\rangle$, colored green in Fig. S9(a), has a fixed eigenvalue $\lambda_B = \lambda_1 = \lambda_2$. When $\gamma_1 < \kappa$, $|\lambda_B\rangle$ actually corresponds to the lossier eigenstate in the *broken phase* and requires the highest gain $g_2 = \kappa^2/\gamma_1$ for oscillation. Thus, it never dominates over the coupled-mode solutions $\{|\lambda_+\rangle, |\lambda_-\rangle\}$ in the symmetric phase. Beyond the EP ($\gamma_1 > \kappa$), however, $|\lambda_B\rangle$ turns into the eigenstate that localizes at the cavity with *gain* in the broken phase and becomes the only stable branch with the lowest oscillation-threshold gain, $g_2 = \kappa^2/\gamma_1 < \gamma_1$.

Finite cavity detuning drastically alters the system response, as shown in Fig. S9(b) and (d) for $\lambda_1 = 1529.72\,\text{nm}$ and $\lambda_2 = 1529.61\,\text{nm}$, which are also used in explaining Fig. 2 (d). In this case, the intense excitation to cavity 2 always couples better with $|\lambda_-\rangle$, because its wavelength $\lambda_-$ is closer to $\lambda_2$ than that of $|\lambda_+\rangle$. Consequently, $|\lambda_-\rangle$ always needs the smallest $g_2$ [Fig. S9(d)] and undergoes the continuous but detuned EP transition without the degeneracy [Fig. S9(b)]. Here, $g_2(\gamma_1)$ for $|\lambda_-\rangle$ has a peaked structure. Thus, if $I_2$ is not large enough to provide the peak threshold gain with $|\lambda_-\rangle$, the system undergoes the suppression and revival of lasing.

We emphasize that the analysis is performed for tracing the revived lasing resonance by $|\lambda_-\rangle$ in Fig. 2(d), and the effect of the detuning is averaged and constant, $2\delta = 0.11\,\text{nm}$. The result is hence approximate and does not explain detailed experimental behavior such as the switching of the dominant mode that is attributed to the pump-induced inversion of $\lambda_1$ and $\lambda_2$.

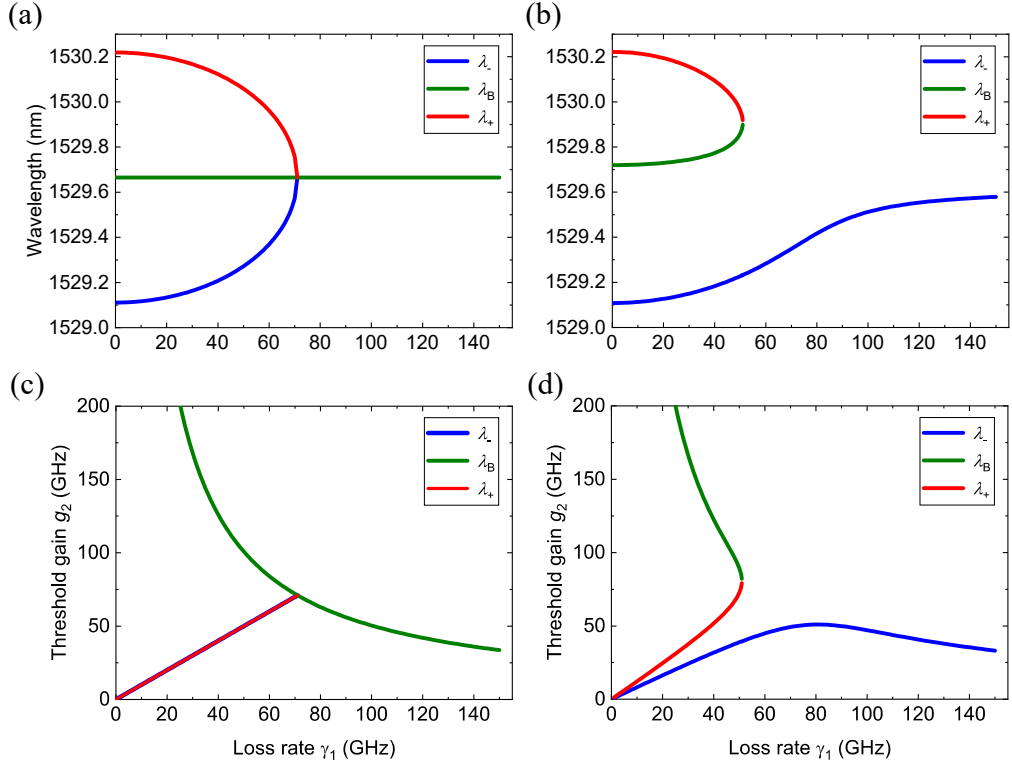

Fig. S9. (a), (b) Steady eigen-wavelengths of the system in the lasing regime dependent on $\gamma_1$, for (a) no cavity resonance detuning $\delta = 0$ and (b) finite detuning $2\delta = 0.11\,\text{nm}$. (c), (d) The corresponding adaptive oscillation-threshold gain $g_2$ for (c) $\delta = 0$ and (d) $2\delta = 0.11\,\text{nm}$. Finite detuning always makes one of the lasing coupled-mode solutions $\{|\lambda_+\rangle, |\lambda_-\rangle\}$ turn into a damping one (not shown), before the system obtains a local loss large enough to reach the EP, $\gamma_{1,\text{EP}} = \kappa = 71\,\text{GHz}$.

The red-side branch $|\lambda_+\rangle$ exhibits peculiar properties in the process. It is stable but inferior compared to $|\lambda_-\rangle$ and thus demands a slightly larger $g_2$, when $\gamma_1$ is small [Fig. S9(d)]. Remarkably, there is an apparent singular coalescence between $|\lambda_+\rangle$ and $|\lambda_B\rangle$ [Fig. S9(b)]. However, these two branches actually denote the *single* eigenstate with different variable gain $g_2$. $|\lambda_+\rangle$ is one of the coupled modes in the symmetric phase. In contrast, $|\lambda_B\rangle$ corresponds to the same eigenstate in the broken phase, which requires the largest $g_2$ by localizing at the lossy cavity. Thus, their coalescence is *not* an EP. The combined solutions for larger $\gamma_1$ come to be a single damping state with $\text{Im}\,\omega_e > 0$, which is hence omitted from Fig. S9(b). This destabilization of $|\lambda_+\rangle$, together with the selective mode excitation by the singularly growing difference of $g_2$ between $|\lambda_+\rangle$ and $|\lambda_-\rangle$, always occurs before reaching $\gamma_{1,\text{EP}} = \kappa$ for any finite $\delta$. Therefore, to achieve the EP of the lasing coupled modes, the detuning is required to be zero over the entire process with a fine accuracy. However, such elaborate control is not realistic in most cases, because there is no direct way to measure the local resonance wavelengths $\{\lambda_1, \lambda_2\}$ for each $\{\gamma_1, \gamma_2\}$ from coupled cavity systems in experiment.

## 11. Selected spectra in lasing regime

The destabilization of $|\lambda_+\rangle$ shown in the last section would result in the unexpected experimental response with multiple peaks split from $|\lambda_+\rangle$ just before the lasing of $|\lambda_-\rangle$. Here, we present and discuss the transition of the emission spectra in the lasing regime $(I_2 = 800\,\mu\text{A})$. When $I_1$ is large [Fig. S10(a)], there are two spectral peaks corresponding to the coupled ground modes with a splitting of about 1.1 nm, which is reasonable for a system with a coupling of $\kappa \approx 60\,\text{GHz}$ and a cavity detuning $\delta$ on an order of 10 GHz. The blue-side mode $|\lambda_-\rangle$ decays faster [Fig. S10(b)] with decreasing $I_1$ and disappears for $I_1 < 30\,\mu\text{A}$. The significant peak power difference between the two modes here is attributed to a non-negligible $\delta$.

The spectral property changes drastically at $I_1 = 5.4\,\mu\text{A}$ [Fig. S10(c)]. The peak power of the red-side mode $|\lambda_+\rangle$ falls by more than 10 dB from the preceding point $(I_1 = 5.6\,\mu\text{A})$, and the device comes to exhibit a weak and broad spectrum spanning from 1529.1 to 1530.2 nm, which is attributed to spontaneous emission. Remarkably, $|\lambda_+\rangle$ is split into two peaks, and the minor one at 1529.9 nm shifts toward the middle of its higher counterpart and the revived $|\lambda_-\rangle$. Note that because cavity 2 is heavily pumped, the amplified spontaneous emission at resonance can have a linewidth below 0.1 nm (or the cold linewidth of each single cavity). Although we also find the first-order mode with an increased intensity (not shown), it never becomes dominant over the ground modes in the entire measurement.

The peak power of the whole spectrum becomes minimum at $I_1 = 2.0\,\mu\text{A}$ [Fig. S10(d)]. Here, $|\lambda_-\rangle$ (left) exceeds the remaining peak around 1530.15 nm (right), which we denote as $|\lambda'_+\rangle$. $|\lambda'_+\rangle$ has an unconventionally spiky structure and hence is not considered as a normal cavity resonance but as a non-steady state [8]. The weaker itinerant bump coming off of $|\lambda_+\rangle$ reaches the right verge of $|\lambda_-\rangle$ and forms a visible plateau. $|\lambda_-\rangle$ is further enhanced and $|\lambda'_+\rangle$ is damped by reducing $I_1$, as shown in Fig. S10(e).

$|\lambda_-\rangle$ is significantly red-shifted and eventually overlaps the shoulder on its recovery [Fig. S10(f)]. Interestingly, while the spectral resolution of the measurement is limited to 0.05 nm, the emission for $I_1 = -5.0\,\mu\text{A}$ (near zero bias of channel 1) fits much better with the squared Lorentzian function than with the Lorentzian. As seen by the localized emission pattern [Fig. 2(e)], the state here is in the broken phase after the detuned EP transition. However, its peculiar spectral shape and position indicate the cancelling of the cavity detuning $\delta$ by reducing $I_1$. It is noteworthy that the measurement equipment used here (OSA) is different from that for the spontaneous emission experiment (a spectrometer).

## 12. Loss-induced lasing in detuned EP transition

The spectral variation with $I_1$ for $I_2 = 800\,\mu\text{A}$ is shown in terms of the peak powers and linewidths of the two major branches in Fig. S11(a) and (b), respectively. The abrupt suppression of $|\lambda_+\rangle$ at $I_1 = 5.4\,\mu\text{A}$ results in the broad spectrum plotted in Fig. S10(c), which includes the re-emergent blue-side peak $|\lambda_-\rangle$ [Fig. S11 (a)]. The power of $|\lambda_-\rangle$ then grows with a kink with decreasing $I_1$, and its linewidth is correspondingly narrowed from $\approx 0.2$ nm to below the minimum measurement resolution here, 0.05 nm [Fig. S11 (b)]. This indicates that the system undergoes the revival of lasing with $|\lambda_-\rangle$ after the switching of the dominant eigenmode.

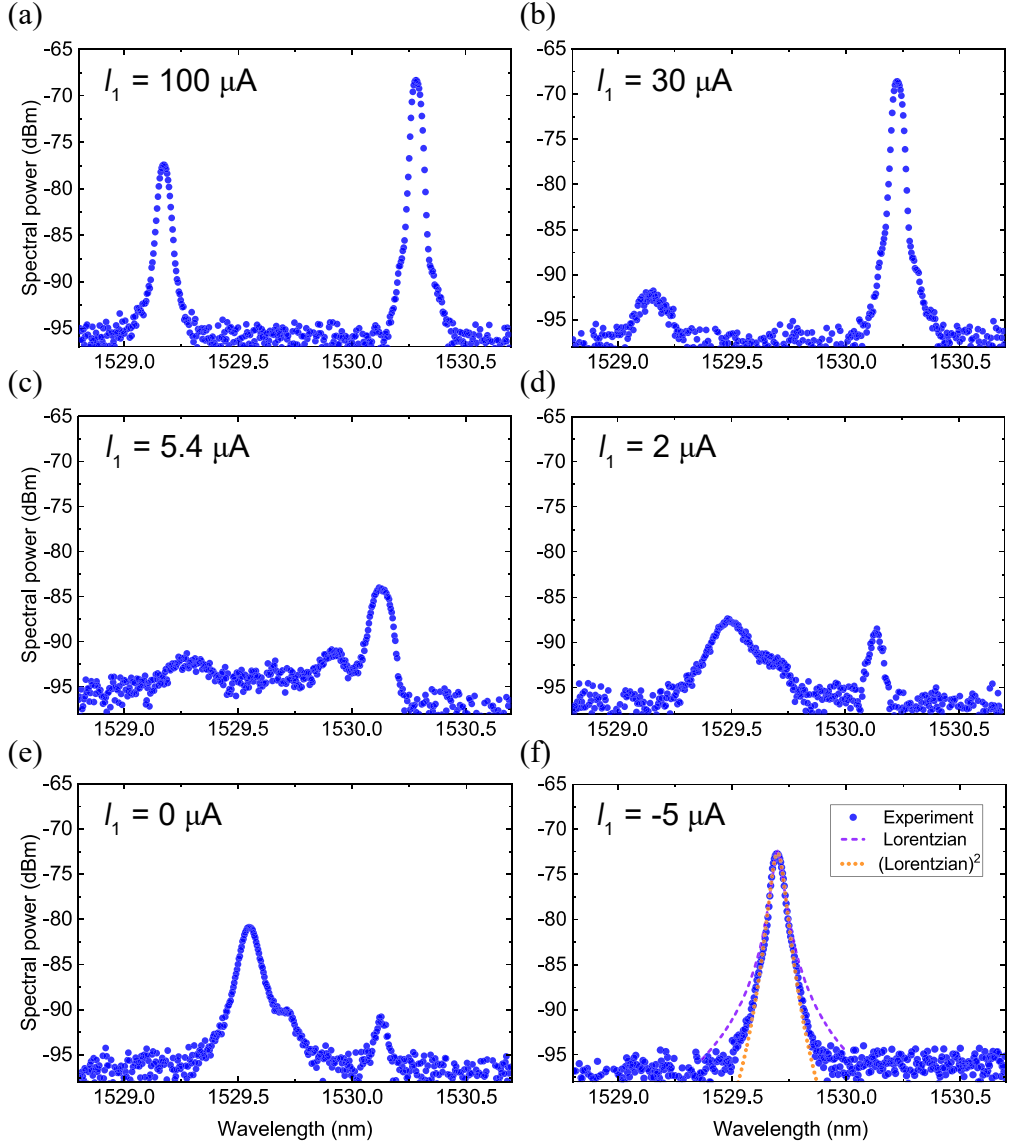

Fig. S10. Coherent emission spectra of the sample under high pumping for cavity 2 ($I_2 = 800\,\mu\text{A}$), for different $I_1$. (a) $I_1 = 100\,\mu\text{A}$, (b) $I_1 = 30\,\mu\text{A}$, (c) $I_1 = 5.4\,\mu\text{A}$, (d) $I_1 = 2.0\,\mu\text{A}$, (e) $I_1 = 0.0\,\mu\text{A}$ and (f) $I_1 = -5.0\,\mu\text{A}$.

It is notable that $|\lambda'_+\rangle$ for $I_1 \leq 5.4\,\mu\text{A}$ denotes the immobile spiky peak remaining around 1530.15 nm, which is again considered unstable [Fig. S10(c)-(e)]. Although it seems to hold somewhat narrow linewidths, it soon decays as $|\lambda_-\rangle$ oscillates. Its detailed properties are beyond the scope of this work.

Finally, the spectra here do not have any hysteresis behavior. They are identical regardless of whether $I_1$ is swept upward or downward, except for a small fluctuation coming from the long-term change in the electrical contact between the device and probes.

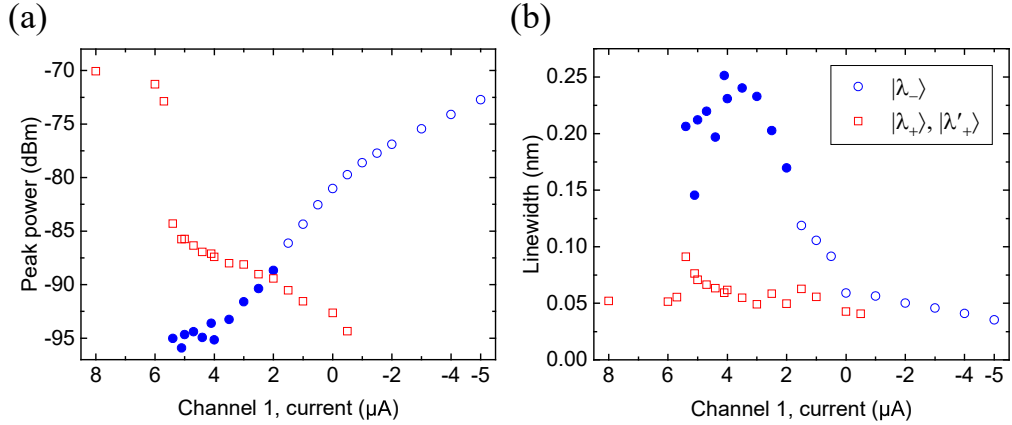

Fig. S11. (a) Peak powers and (b) linewidths of the $|\lambda_+\rangle$, $|\lambda'_+\rangle$ (red) and $|\lambda_-\rangle$ (blue) under small $I_1$. Filled blue markers: approximate linewidths estimated with the moving-average data. The plots for $|\lambda'_+\rangle$ after the drop of the power correspond to the immobile peak around 1530.15 nm in Fig. S9(c)-(e), which is attributed to the non-steady damping state. Another weaker and broader bump moving toward the middle is omitted. The peak power rising and linewidth narrowing of $|\lambda_-\rangle$ indicates the revival of lasing. Here, the minimum linewidth resolution is 0.05 nm.



\end{document}